\documentclass[journal]{IEEEtran}
\usepackage[dvipsnames]{xcolor}

\usepackage{cite}
\usepackage{array}
\usepackage{amsmath,amssymb,amsfonts}
\usepackage{textcomp}
\usepackage{svg}
\usepackage{booktabs}
\usepackage{multirow}
\usepackage{graphicx}
\usepackage{subcaption}
\usepackage[linesnumbered,ruled,vlined]{algorithm2e}
\usepackage{algpseudocode}
\usepackage{threeparttable}
\usepackage{orcidlink}
\usepackage{caption}
\usepackage{graphicx}

\ifCLASSINFOpdf
\else
\fi

\begin{document}
\setlength{\dbltextfloatsep}{0pt}
\title{Dynamic Tsetlin Machine Accelerators\\ for On-Chip Training using FPGAs}

\author{Gang Mao$^{\orcidlink{0009-0009-2535-7149}*\dagger}$, Tousif Rahman$^{\orcidlink{0000-0001-8669-010X}*\dagger}$, Sidharth Maheshwari$^{\orcidlink{0000-0002-9665-5698}*\ddagger}$, Bob Pattison$^{\orcidlink{0009-0001-2822-8408}\dagger}$, \\Zhuang Shao$^{\orcidlink{0000-0001-7824-0985
 }\dagger}$, Rishad Shafik$^{\orcidlink{0000-0001-5444-537X}\dagger}$, \textit{Senior Member}, Alex Yakovlev$^{\orcidlink{0000-0003-0826-9330
}\dagger}$, \textit{Fellow} \thanks{This work was supported by EPSRC EP/X036006/1 Scalability Oriented Novel Network of Event Triggered Systems (SONNETS) project and by EPSRC EP/X039943/1 UKRI-RCN: Exploiting the dynamics of self-timed machine learning hardware (ESTEEM) project.} \thanks{$^*$Indicates equal contribution.} \thanks{$^\dagger$Microsystems Research Group, School of Engineering, Newcastle University, Newcastle upon Tyne, NE1 7RU, United Kingdom (UK).} \thanks{$^\ddagger$IIT Jammu, NH-44 , PO Nagrota, Jagti, Jammu \& Kashmir 181221, India.}}

\markboth{Preprint - Accepted in IEEE Transactions on CIRCUITS AND SYSTEMS—I: REGULAR PAPERS}%
{Mao and Rahman \MakeLowercase{\textit{et al.}}: Dynamic Tsetlin Machine Accelerators for On-Chip Training at the Edge using FPGAs}

\maketitle

\begin{abstract}
The increased demand for data privacy and security in machine learning (ML) applications has put impetus on effective edge training on Internet-of-Things (IoT) nodes. Edge training aims to leverage speed, energy efficiency and adaptability within the resource constraints of the nodes. Deploying and training Deep Neural Networks (DNNs)-based models at the edge, although accurate, posit significant challenges from the back-propagation algorithm's complexity, bit precision trade-offs, and heterogeneity of DNN layers. This paper presents a Dynamic Tsetlin Machine (DTM) training accelerator as an alternative to DNN implementations. DTM utilizes logic-based on-chip inference with finite-state automata-driven learning within the same Field Programmable Gate Array (FPGA) package. Underpinned on the Vanilla and Coalesced Tsetlin Machine algorithms, the dynamic aspect of the accelerator design allows for a run-time reconfiguration targeting different datasets, model architectures, and model sizes without resynthesis. This makes the DTM suitable for targeting multivariate sensor-based edge tasks. Compared to DNNs, DTM trains with fewer multiply-accumulates, devoid of derivative computation. It is a data-centric ML algorithm that learns by aligning Tsetlin automata with input data to form logical propositions enabling efficient Look-up-Table (LUT) mapping and frugal Block RAM usage in FPGA training implementations. The proposed accelerator offers 2.54x more Giga operations per second per Watt (GOP/s per W) and uses 6x less power than the next-best comparable design.

\end{abstract}

\IEEEpeerreviewmaketitle

\section{Introduction}
\IEEEPARstart{M}{achine} Learning (ML) offers a generalized approach to developing autonomous applications from "Internet-of-Things" (IoT) sensor data. Having ML execution units in close proximity to the sensor, at the so-called \textit{edge}, enables faster task execution with high data security and privacy. However, sensor degradation and environmental factors may require recalibration~\cite{Recalibration} or user-personalized on-field \textit{training}~\cite{Personalized_Learning} to ensure continued functionality. Implementing solutions to these challenges is nontrivial. It requires finding the right balance between achieving the appropriate learning efficacy for the ML problem and the restrictive compute\slash memory resources available on the platforms~\cite{survey}. 

For ML \textit{inference} tasks on edge nodes, these challenges have been widely explored, e.g., quantization~\cite{FINN_R}, sparsity-based compression, and pruning for the most commonly used Deep Neural Network (DNN) models~\cite{survey, FINN, Zhang2017}. So far, very few works have dealt with DNN on-field \textit{training} due to two important design challenges: firstly, the main compute for DNN on-field training typically requires retaining full precision floating point general matrix multiplications and secondly the complexity of the backpropagation algorithm through the heterogeneous DNN layers~\cite{chen2023}. These challenges materialize from algorithmically intrinsic attributes of DNNs. State-of-the-art approaches to alleviating these challenges are explored in Section~\ref{sec:related_works}. This work addresses the aforementioned challenges in on-field training through two fronts: firstly, on the algorithm front, whether the same (or better) edge training performance can be harnessed from simpler and more hardware-oriented ML algorithms; and secondly, on the hardware front, how such algorithms can be efficiently mapped to edge training platforms. These fronts are addressed in turn:

On the algorithm front, this work explores ML algorithms with lower training complexity compared to DNNs called Tsetlin Machines (TMs)~\cite{granmo2021tsetlinmachinegame}. The TM and its variants are logic-based learning algorithms. They learn logical propositions called \textit{clauses} to identify classes. The learning process for creating these logic propositions involves state transitions of finite-state automata. The TM and its family of algorithms (\hspace{1sp}\cite{CoalescedTM, ConvTM, regressionTM}) alleviate the above two core challenges faced by DNNs: TMs utilize logic-based computation and simpler nested decision logic to adjust the learning elements during training instead of floating-point weights and backpropagation.  

On the hardware front, this work argues the case for very resource-constrained, low-cost, low-power, System on Chips (SoCs) operating as on-site single edge nodes. The recent adoption of such SoCs as embedded FPGAs (eFPGAs) offers the ability to integrate custom hardware accelerators in conjunction with CPU processing to leverage faster inference latency within edge node power budgets~\cite{chen2023, Zhang2017}. The FPGA component, in particular, offers advantages in customization, cost, ease of prototyping, and lower implementation risk compared to ASIC solutions with better power efficiency than CPU and GPU training implementations~\cite{Edge_Compute_survey}. Additionally, FPGA accelerator designs can be easily scaled to smaller or larger platforms, depending on the application needs. The culmination of the two fronts above forms the main contributions towards the development of Dynamic Tsetlin Machine (DTM) FPGA hardware implementations:  

\begin{figure*}[t]
    \centering
    \includegraphics[width =0.89\linewidth]{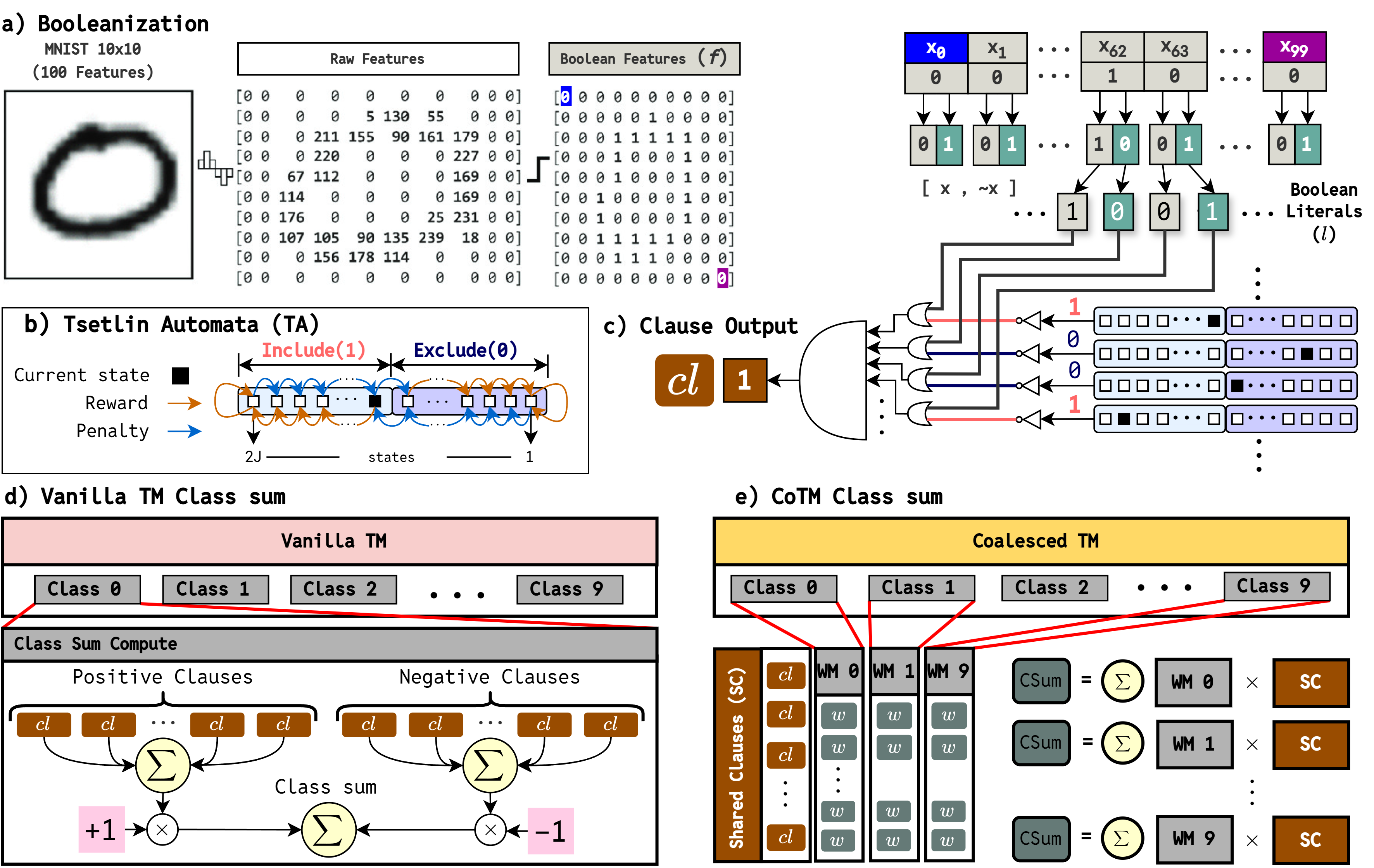}
    \caption{\small{Block diagram of the fundamental components of the Tsetlin Machine algorithms for \textit{inference} - they build the Vanilla TM (pink) and the Coalesced TM (CoTM) (yellow). Component breakdown: a) shows the pre-processing or \textit{Booleanization} process (adapted from~\cite{Olga_reproduciable}) to generate the inputs to TMs using an MNIST datapoint example; b) shows the learning element - the \textit{Tsetlin Automata (TA)}; c) shows how each TA relates to its respective Boolean literal to create the \textit{Clause Output}. The pink and yellow blocks for the Vanilla and CoTM show \textit{Class sum} computation for each class (d and e).}}
    \label{fig:CoTM_block}
\end{figure*}
\begin{figure*}[t]
    \centering
    \includegraphics[width =0.97\linewidth]{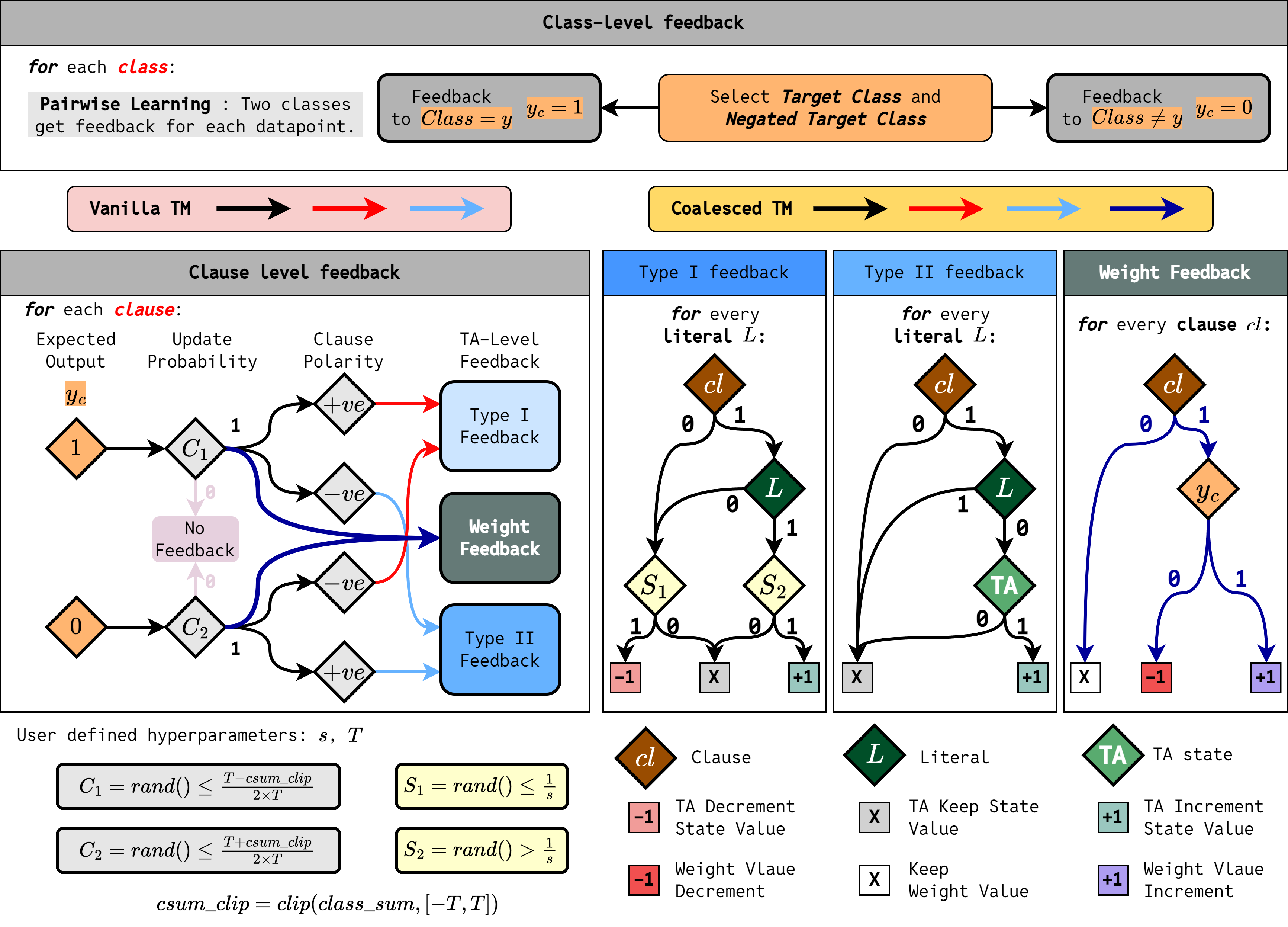}
    \vspace{-3mm}
    \caption{\small{Block diagram of \textit{learning} process used in the Vanilla TM and CoTM referred to as \textit{feedback}. Feedback is used to transition the learning elements of the model, i.e., the TAs and weights (weights for CoTM). These processes are presented visually here, but will be explored as algorithm blocks in the subsequent section.}}
    \label{fig:TM_feedback}
\end{figure*}

\begin{itemize}
    \item DTM envisages generalized FPGA hardware for training and inference for all variants of TM algorithm. This paper currently generalizes \textit{two} variants of the TM algorithm viz. Vanilla and Coalesced TM (CoTM)~\footnote{The standard TM algorithm as described in~\cite{granmo2021tsetlinmachinegame} will be referred to as Vanilla but still abbreviated as TM.}.
    \item \textbf{Pseudo Random Number (PRN) Bandwidth and quality:} Efficient TM training requires a large amount of high-quality PRNs with high bandwidth. Considering resource constraints at the edge, an XOR-shift-based re-seeding mechanism is developed providing fresh seeds after 2L cycles for every L-bit Linear Feedback Shift Registers (LFSRs) enhancing training accuracy for small L.
    \item \textbf{Scalability: }The DTM architecture can be scaled according to the available hardware resource budget but retains the capability to execute a range of model sizes. 
    \item \textbf{Flexibility:} For an implemented hardware architecture, DTM enables data-centric inter-model switching through precomputed masks and matrix iteration cycles at run-time without needing resynthesis (Section IV-A). 
    \item \textbf{Reducing computation complexity:} DTM training improves design metrics by replacing floating-point arithmetic with integer-arithmetic retaining desired accuracy (Section IV-B). 
    \item \textbf{Optimized Clause-Level feedback:} Feedback to clauses reduces as the model converges. DTM optimally skips groups of clauses that do not receive any feedback in the TA update stage, therefore, reducing the total number of memory accesses and operations (Section IV-B).
\end{itemize}

The remaining sections explore these contributions through examining TM algorithms and related designs in Sections II and III, the proposed architecture in Section IV and evaluation in Section V. 
   
\vspace{-2mm}
\section{Background: Tsetlin Machine Algorithms}
This paper introduces the first CoTM hardware and Dynamic TM architecture. This section presents the fundamental components of the algorithms in the DTM architecture.

\textbf{\textit{A. Inference:}} Figure \ref{fig:CoTM_block} shows inference components in four parts: a) shows input data preprocessing called \textit{Booleanization}, b) shows \textit{Tsetlin Automata} (TA) - the learning element in TM algorithms,  c) shows how processed input data interact with their respective TAs to create a \textit{Clause Output}, and the two Vanilla and CoTM blocks show how class sums are computed. 

\textbf{a) Booleanization:} Booleanization is the process of converting raw input data into a binary encoding and extending this to each binary feature and complement ($x, \tiny{\sim}{x}$) - these are referred to as Boolean features (seen in grey) and Boolean Literals (\texttt{\textit{l}}) (grey and green), respectively, as seen in Fig.~\ref{fig:CoTM_block}a with an MNIST~\cite{6296535} \texttt{0} digit example. In this example, only one threshold is used to convert the raw feature to the Boolean Feature (\texttt{\textit{f}}). However, in practice, many thresholds can be applied to each raw feature such that they are binned into discrete binary encodings. Details on Booleanization and different Booleanization strategies are explored in detail in~\cite{Booleanization}. Higher complexity datasets such as CIFAR-10 and CIFAR-100 are targeted using Composite TM architectures~\cite{grønningsæter2024optimizedtoolboxadvancedimage, granmo2023tmcompositeTM}. Composite TMs involve a \textit{group} composed predominantly of Convolution TMs~\cite{ConvTM} that each work on their own respective Booleanized input space derived via adaptive thresholding (adaptive Gaussian and Otsu with kernels up to $10 \times10$), Canny edge detection, Color thermometers (kernels of $3 \times3$ to $5 \times5$) and Histogram of Gradients. Vanilla and Coalesced TMs feature less prevalently in such Composite TM architectures being suitable for smaller multivariate, image and audio problems (explored later in Section V). Future work will add the Convolutional TM modules and pre-processing frameworks as modules for DTM enabling Composite TM training support. 

\textbf{b) Tsetlin Automata (TA):} The fundamental component underpinning all TM algorithms is Tsetlin Automata (TA). Each TA behaves like a finite state machine with two possible actions with respect to its corresponding Boolean literal. As seen in Fig. 1b, TA has 2J states; if the automaton’s state value is $\leq$ J, then the automaton’s action is to Exclude the corresponding literal, while if the state value $>$ J, then the automaton’s action is to Include the corresponding literal (Fig.~\ref{fig:CoTM_block}b). These Include/Exclude decisions allow the TM to build logic expressions seen in the next image section (Fig.~\ref{fig:CoTM_block}c). 

The convergence of TMs relies on the self-organizing behavior of TAs as they learn optimal logic expressions to classify data. Each TA operates independently, receiving Reward (brown arrow) and Penalty (blue arrow) signals to transition their state as they compete to form the best decision boundaries, their interactions can be modeled as a game where each automaton tries to maximize classification accuracy. A Nash Equilibrium is reached when no TA can improve its decision without detriment in the performance of another~\cite{Convergence_TM}. 

\textbf{c) Clause Output:} Fig.~\ref{fig:CoTM_block}c shows how each Boolean literal in a MNIST datapoint is interfaced with its own TA to create a clause. The clause computation is composed of \texttt{AND, OR} and \texttt{NOT} operations and generates a single bit \textit{Clause Output} ($cl$ marked with brown). The training process determines which Boolean literals to include and exclude using their respective TAs to build logic expressions for each clause. Clauses are the fundamental modules of TMs; however, they are used differently in each TM algorithm. The pink and yellow boxes illustrate clause usage for class sums in Vanilla and CoTMs.

\textbf{d) Class sum compute for Vanilla TM:} Focusing on Vanilla TM first, Fig.~\ref{fig:CoTM_block}d shows how each class is composed of a set number of clauses. The clauses are divided into two teams (Positive Team and Negative Team). The one-bit outputs of each clause are summed and multiplied by either $+1$ or $-1$ polarity depending on the clause team. The summation of both teams' polarity-adjusted sums produces the class sum. The argmax of each of these class sums across all classes generates the final inference classification.

\begin{figure}[t]
    \centering
    \includegraphics[width =\linewidth]{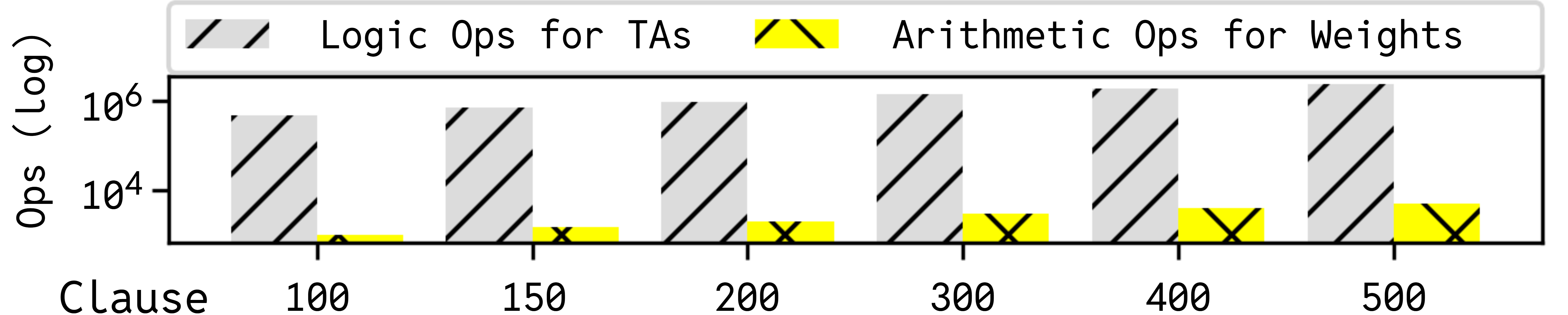}
    \vspace{-5mm}
    \caption{\small{The log scale comparison of logic-based (clause compute) vs. integer-based (class sum compute) arithmetic operations in CoTM inference. The number of operations increase with clauses.}} 
    \label{fig:CoTM_Scale}
    \vspace{-6.5mm}
\end{figure}

\textbf{e) Class sum compute for CoTM:} CoTM uses clause outputs differently. Instead of instancing clauses for every class, it creates a shared pool of clauses (SC), shown in Fig.~\ref{fig:CoTM_block}e. These clause outputs are shared across all classes. To generate a class sum, each class has its own set of learnable \textit{signed integer} weights ($w$) which act as multiplicands with their respective clause in the SC pool. The clauses in the SC pool no longer have polarity like the Vanilla TM; instead, their polarity is assigned when multiplied with their respective weight in each class. Despite the additional weight multiplications, as seen in Fig.~\ref{fig:CoTM_Scale}, the inference operations are still dominated by the logical operator computation.  

\textbf{\textit{B. Learning (Feedback):}} \textbf{a) TM learning algorithms (feedback):} During training, the learning process occurs for each datapoint after the inference is completed, i.e. the clause outputs and class sums have been generated. TM learning algorithms find the optimum states for every TA in the system and the values for the weights for each clause for every class of the CoTM. To transition a TA, that is, increment or decrement its state, \textit{feedback} must be issued down the hierarchy of the TM. Fig.~\ref{fig:TM_feedback} shows how this first starts with Class-level feedback, to Clause-level feedback, and finally how this leads to TA increment and decrements.

\textbf{b) Class level feedback:} Feedback decisions for both TM algorithms begin at the class level, and both Vanilla and CoTM follow the same class-level feedback procedure. The learning in pairwise, for each datapoint, two rounds of feedback will happen (these are referred to as class updates), each round updates one particular class. Firstly, the target class $Class = y$ (that is, the class to which the data point corresponds) followed by a randomly chosen class $Class \neq y$. If the target class is selected, then $y_c = 1$ otherwise $y_c = 0$. This forms the inputs to the decision process used for clause-level feedback. 

\textbf{c) Clause level feedback:} Once the class $y_c$ has been selected for feedback, the class sum of this chosen class will be used to evaluate the clause update probability $C_1$ or $C_2$ (the class sum would already have been computed during inference). The clause update probability is a comparison between a random number and an expression that is formed by the clipped class sum ($csum\_clip$) and the \textit{Threshold} hyperparameter $T$. If the clause update probability is $1$ then this clause has been selected for feedback. For Vanilla TM, the selected clause's polarity is used to determine whether it should receive Type I or Type II feedback. These are used to provide feedback at the TA level. 

\textbf{d) Clause level feedback (Weight Feedback):} As seen in the inference figure, CoTM has a weight assigned to each clause in each class. The weight feedback is dependent on the generated clause output for this particular clause and the class-level feedback $y_c$. The weight update is controlled through the clause update probabilities $C_1$ or $C_2$. 

\textbf{e) TA level feedback:} Type I and Type II feedbacks both iterate through every literal in the selected clause to determine whether to transition it's associated TA. For Type I feedback, there are TA update probabilities $\frac{1}{s}$ and $\frac{s - 1}{s}$ for both TA state value increment and decrement, respectively. These update probability expressions introduce the \textit{Sensitivity} hyperparameter $s$. TM algorithms also have a mode that is called "boost true positive". When using this mode, the increase in Type I feedback will always occur regardless of the condition $S_2$. Type II feedback is deterministic. The feedback process involves two hyperparameters $s$ and $T$. The effect of these parameters on TA updates and clause updates, respectively, is explored in~\cite{Olga} for the Vanilla TM. The following section will illustrate how these probabilities are translated to FPGA. 

\vspace{-2mm}
\section{Related Accelerator Designs}
\label{sec:related_works}

This section explores training accelerator designs targeting FPGAs for DNNs to address how intrinsic attributes of the TM algorithm can offer reduced complexity\footnote{The related works are qualitatively compared here to understand the design choices. In the Accelerator Evaluation section, these implementations are considered more quantitatively against the proposed accelerator.}.  

\textbf{a) Quantization in Training:} One of the algorithmic drawbacks of all DNN-based algorithms is the use of 32-bit floating point weights and computation. Most works address this through \textit{quantization} of the these floats~\cite{FINN_R, FP_BNN, FracBNN, CNN_1, YOLOv3, A2NN, WSQ, 8587697}. Extreme quantization to binary weights and data, referred to as Binary Neural Networks (BNNs), leads to a reduction in the computation of floating point multiplication and addition to \texttt{XNOR} and \texttt{popcount}. This is incredibly effective for high-throughput inference accelerators~\cite{FINN_R, FP_BNN, FracBNN}. Translating the \textit{training} of these quantized DNNs to accelerators is challenging because the precision of the weights cannot be reduced as significantly. For example, for training, the authors in~\cite{CNN_1} retain 20-bit weights for their accelerator. This results in substantial Block RAM (BRAM) usage for weight storage and many DSP blocks for their computation. Batched training to improve throughput is seen in~\cite{reconfigurable_dnn} (Low Batch CNN).

 \textbf{b) Design Approaches for Backpropagation:} DNN training requires backpropagation for weight updates during training. This involves computationally expensive partial derivative calculations. The backpropagation computation in the training accelerators in~\cite{DNN_1}, is kept as floating point, but happens on the ARM processor of an SoC-FPGA. The proposed accelerator is also developed for an SoC-FPGA but the ARM processor is used for configuring the run-time architecture search. The work in~\cite{reconfigurable_dnn} (Reconfig DNN) also provides reconfigurability to adopt different applications.

 \textbf{c) Addressing Layer Heterogeneity: }Stochastic gradient descent allows training in minibatches. This approach for backpropagation is used by~\cite{9256704, F_CNN, BFP-NN, FlexBlock}, but this time developed on the FPGA fabric itself; this approach retains 32-bit floats throughput. This work also addresses another design challenge with DNNs: Layer heterogeneity. They partition the training process according to the different layers and use a stream-based reconfiguration to configure basic modules for each layer's workload. The main overhead is memory transfer for each layer from off-chip memory.     

 \textbf{d) Spiking Neural Network Approaches: }Two digital Spiking Neural Network (SNN) inference FPGA implementations are reported in~\cite{SNN_1, FireFly}. For the three-layer SNN network~\cite{SNN_1} mapped to Xilinx Kintex-7 reported in this paper, it can infer $233$ images per second for MNIST and FMNIST, while achieving $97.81\%$ and $83.16\%$ accuracy. In~\cite{SATA}, a 65nm SNN training AISC simulation is reported. The simulated training throughput for MNIST is around 5000 images/s with $99.00\%$ test accuracy. 
 \begin{figure*}[ht]
    \centering
    \includegraphics[width = 0.95\linewidth]{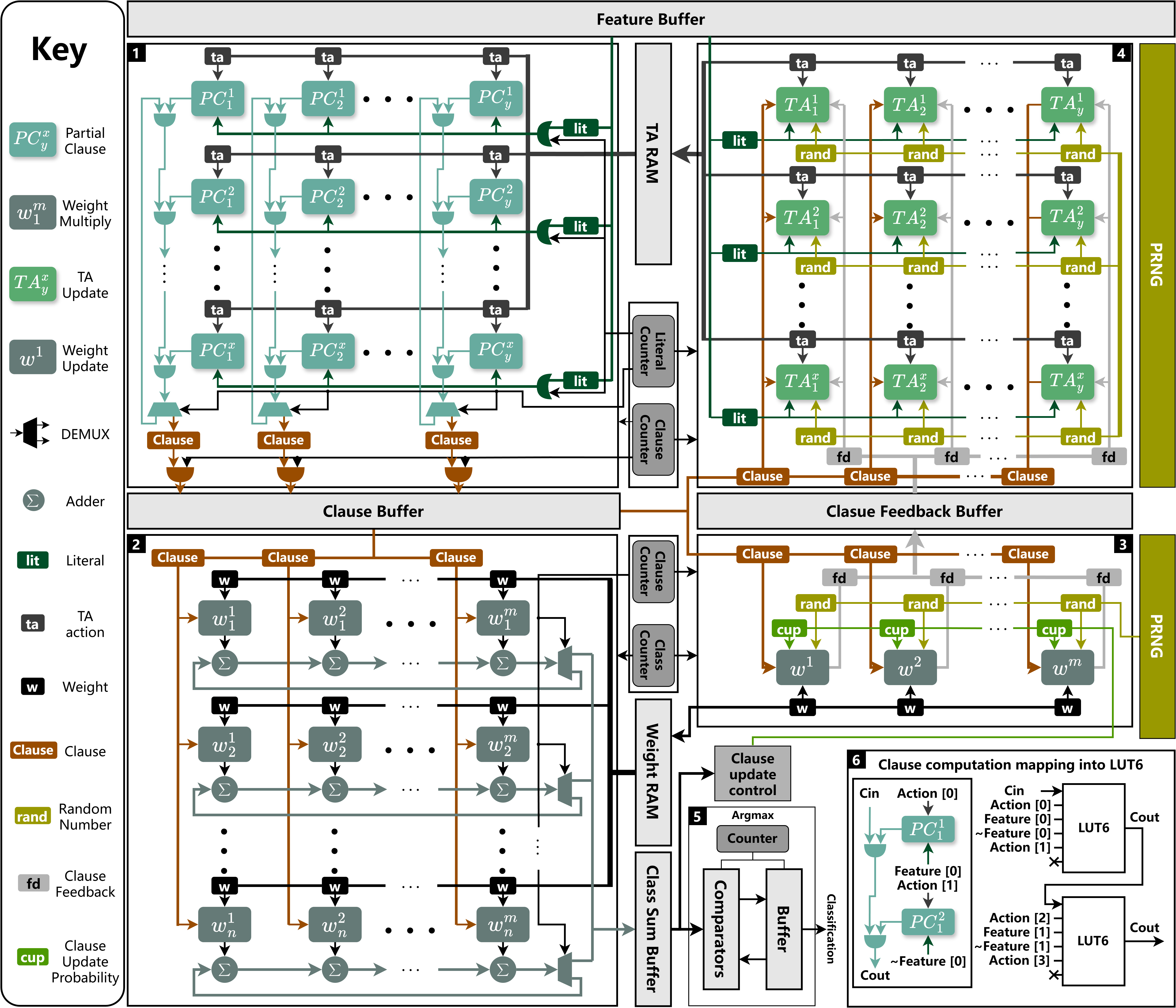}
    \vspace{-1mm}
    \caption{\small{The architecture of the proposed \textbf{Dynamic Tsetlin Machine} (DTM) training accelerator. The DTM accommodates both the Vanilla and CoTM algorithms. It consists of 5 core modules: \textbf{1}-Clause Matrix, \textbf{2}-Weight Matrix, \textbf{3}-Weight Update Matrix,  \textbf{4}-TA State Update Matrix, \textbf{5}-Argmax Output.} Blocks \textbf{1}, \textbf{2} and \textbf{5} are for Vanilla and CoTM inference while TM training needs both blocks \textbf{3} and \textbf{4}. All buffers and Pseudo Random Number Generator (PRNG) blocks are used for both algorithms. \textbf{6} also shows how the clause computation is mapped into LUT6 FPGA elements.}
    \label{fig:CoTM_Arch}
\end{figure*}

\textbf{e) Advantages of TMs and related works: } The TM algorithms intrinsically use bitwise operations for clause computation and comparisons using random numbers in the feedback process. The entire dataflow can, naturally, operate in the integer-arithmetic domain on hardware mitigating many design challenges encountered in quantization and backpropagation. The first preliminary TM training and inference implementation~\cite{RoySoC} uses an ultra-low power ASIC solution based on 65nm technology customized for the 3-class Binary Iris data set~\cite{misc_iris_53}. REDRESS~\cite{REDRESS} presents a microcontroller-based custom high-throughput inference implementation employing bit-parallel optimization. A custom TM accelerator framework (MATADOR)~\cite{Matador} focuses on automated deployment for maximum throughput, once again for a specific task, with a \textit{custom hard-wired} Vanilla TM model. In MATADOR, the TA actions are hardcoded into LUTs and cannot be calibrated without resynthesis, which means MATADOR cannot be extended to a training accelerator. A recent FPGA-based customized Convolutional Tsetlin Machine (Conv TM) training accelerator~\cite{Svein_Conv} reports solving customized $28 \times 28$ booleanized images, with a $10 \times 10$ convolution window. Conv TM is designed for fast throughput at the expense of resources. Section~\ref{sec:result} presents further comparison between DTM and Conv TM. In contrast to related works, the DTM design showcases real-time recalibration and flexibility at the expense of throughput. These qualities are more valuable in on-chip training implementations that require re-calibration.

\section{Proposed Dynamic Tsetlin Machine (DTM) Accelerator}
\label{sec:architecture}

\begin{figure*}[t]
    \centering
    \begin{subfigure}[b]{0.45\textwidth}
        \includegraphics[width=\textwidth]{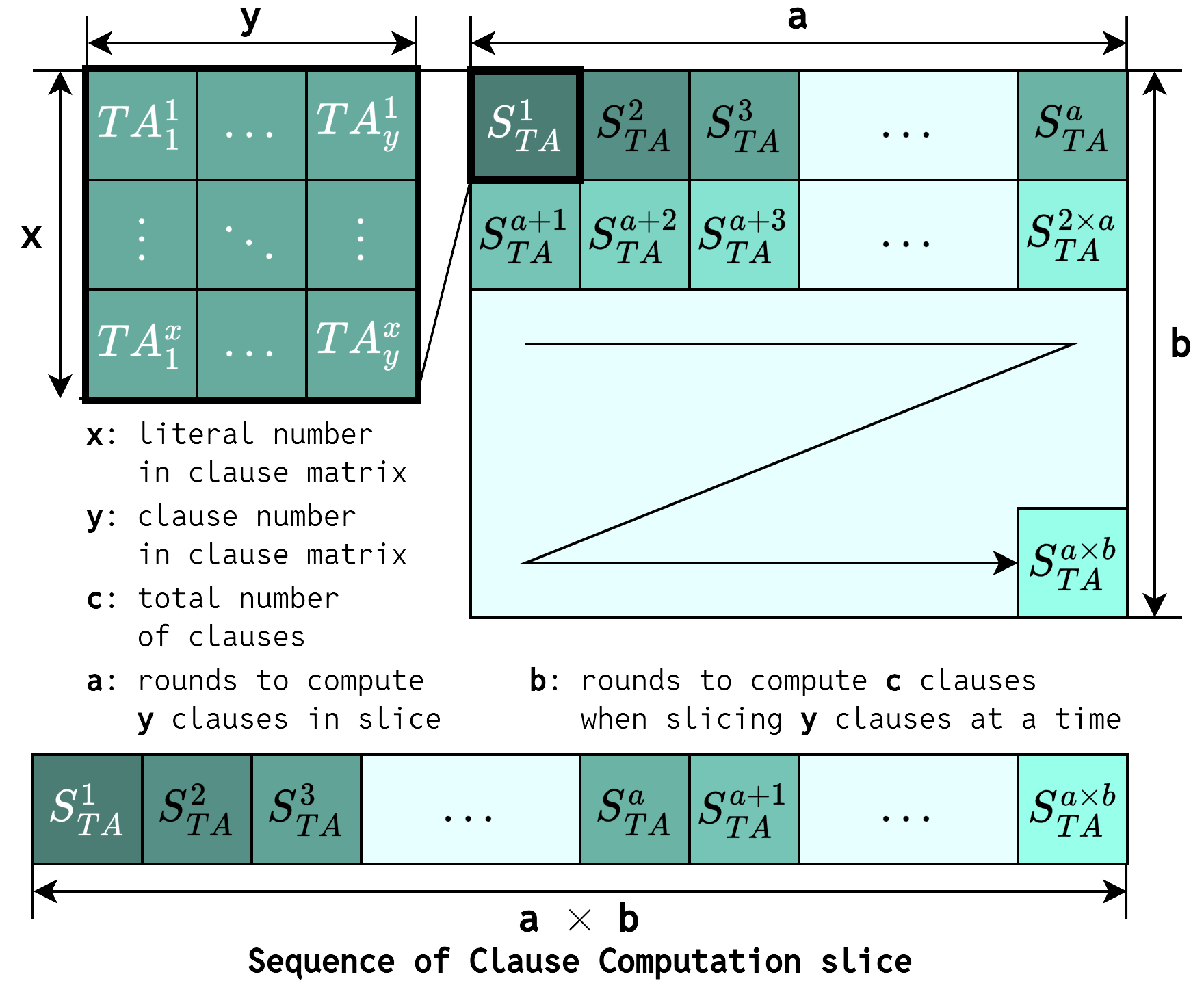}
        \vspace{-6mm}
        \caption{Partial clause computation slices and iteration sequence ($S_{TA}$ represents one round of compute for the \texttt{Clause Matrix}).}
        \label{fig:mem_clause_slice}
    \end{subfigure}
    \begin{subfigure}[b]{0.45\textwidth}
        \includegraphics[width=\textwidth]{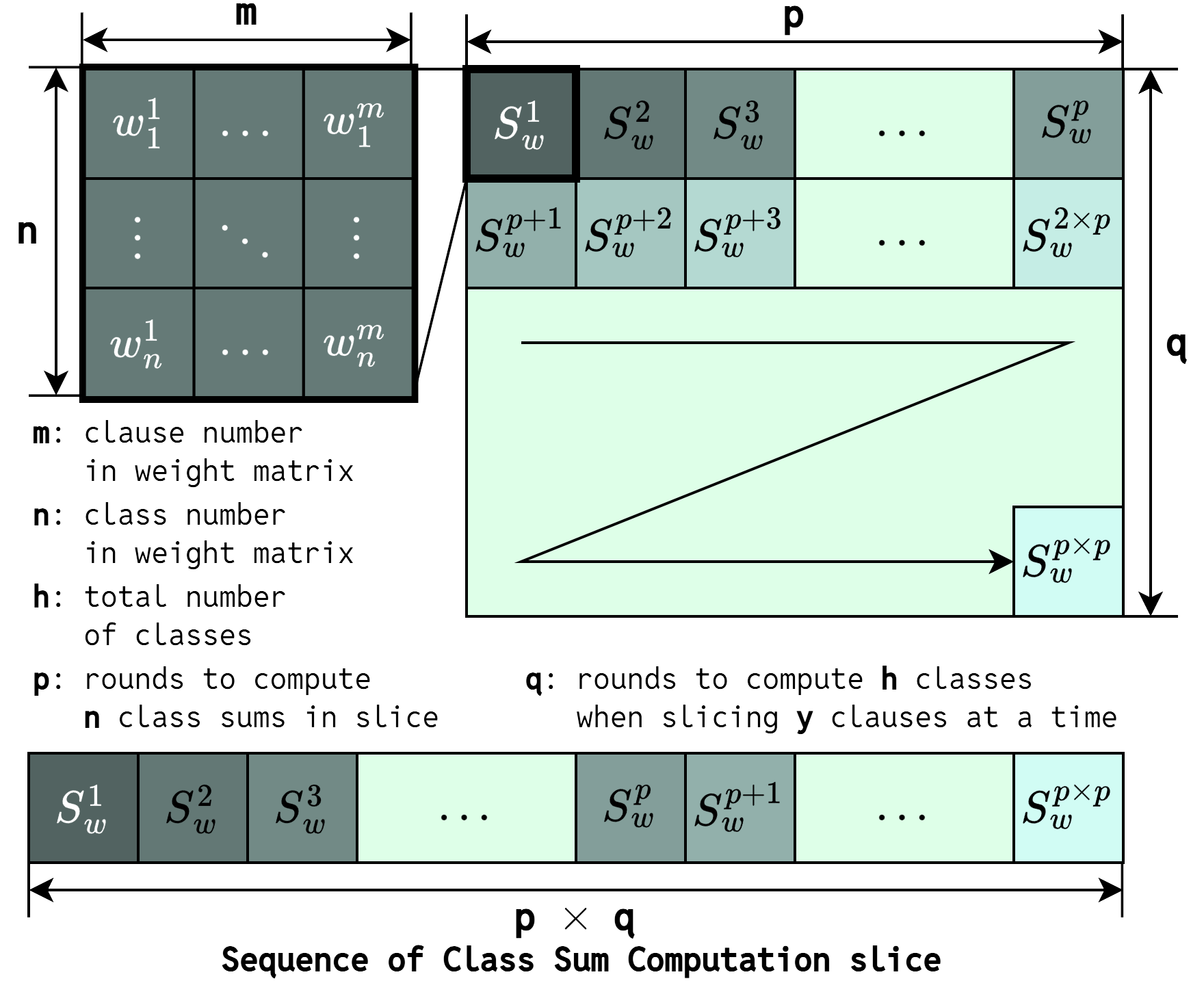}
        \vspace{-6mm}
        \caption{(Class sum computation slice in memory ($S_{w}$ represents the {Weight Matrix} computation slice).}
        \label{fig:mem_weight_slice}
    \end{subfigure}
    \vspace{-1.5mm}
    \caption{\small{Visualizing the iterations required to compute full clause and weight matrices. These figures show the rounds of compute required for the \texttt{Clause Matrix} and \texttt{Weight Matrix}. The colors of (a) and (b) correspond to the colors of the matrix modules in Fig.~\ref{fig:CoTM_Arch}.}}
    \label{fig:Mem_slice}
\end{figure*}

The following two subsections explore how the TM and CoTM concepts are translated to the DTM architecture implementation with SoC integration. Fig.~\ref{fig:CoTM_block} shows that the TM algorithms are intrinsically modular with the Vanilla and CoTM architectures composed of clause computation blocks. Unlike layered DNN architectures, the model architectures of both Vanilla and CoTMs are controlled by the number of clauses (alleviating layer heterogeneity design challenges). This shared basis across the two TM algorithms allows for greater reuse. The \textit{Dynamic} in the DTM architecture represents flexibility and reconfigurability at runtime.

The compute and memory components of the DTM architecture are presented in Fig.~\ref{fig:CoTM_Arch}. The architecture is divided into five compute modules: \texttt{Clause Matrix}, \texttt{Weight Matrix}, \texttt{Weight Update Matrix}, \texttt{TA Update Matrix} and \texttt{Argmax} and their associated memory modules. The \texttt{Clause Matrix}, \texttt{Weight Matrix} and \texttt{Argmax} are used for inference, while the remaining update blocks are used for training. The next two subsections examine these blocks and their data buffers in inference and training.


\textbf{\textit{A. DTM Inference Modules:}}
\label{subsec:inference}
Inference involves computing of clause outputs (for every class for Vanilla TM), applying the weights (for the CoTM) and then summing (class sums) and performing argmax. This subsection walks through this process from how features enter the accelerator to how the classification is generated. 

\textbf{a) Feeding in the Features:}
\label{subsubsec:feature buffer}
The accelerator is interfaced with the SoC processor via an AXI4-Stream channel, Boolean input features are received and placed in the \texttt{Feature Buffer} (see Fig.~\ref{fig:CoTM_Arch}). Users does not need to set this buffer to the problem's feature size explicitly, the feature buffer size only determines a maximum feature number. If the DTM accelerator is to be deployed on a resource-limited FPGA chip (e.g., Zynq7020) the \texttt{Feature Buffer} will be registers, but when scaling to larger chips (e.g., Ultrascale + ZU-7EV), the \texttt{Feature Buffer} will be instanced as BRAM. This allows DTM to handle larger datasets in a more resource-efficient way. Suppose that the width of the AXI bus is \(W\) bits; the DTM will receive \(W\) Boolean features in every clock cycle. In the clause computation phase, it selects \(\frac{x}{2}\) features and generates their respective \(x\) literals (these features and their complements) and sends them to the \texttt{Clause Matrix}. 

\textbf{b) Clause Computation:}
\label{subsec:Clause_matrix}
Clause computation occurs in the \texttt{Clause Matrix} (Block \textbf{1} in Fig.~\ref{fig:CoTM_Arch}), the detailed mapping of the clause computation to the LUTs is shown in Fig.~\ref{fig:CoTM_Arch}-\textbf{6}. Loading all literals (\texttt{lit}) and TA states (\texttt{ta}) for every TM clause at once demands high bandwidth and logic resources. Complete clause computations create long combinational chains, reducing operational frequency; to reduce resource pressure and latency, TA actions won't all load into the \texttt{Clause Matrix} at once.

\vspace{-1mm}
\begin{equation} \label{eq:clause_compute_matrix}
\begin{bmatrix}
    p\_cl_{1}\\
    p\_cl_{2}\\
    \vdots\\
    p\_cl_{y}\\
\end{bmatrix}
\land
\begin{bmatrix}
    \land_{i = 1}^{x} (L_{i} \lor \sim TA^{i}_{1})\\
    \land_{i = 1}^{x} (L_{i} \lor \sim TA^{i}_{2})\\
    \vdots\\
    \land_{i = 1}^{x} (L_{i} \lor \sim TA^{i}_{y})\\
\end{bmatrix}
=
\begin{bmatrix}
    p\_cl'_{1}\\
    p\_cl'_{2}\\
    \vdots\\
    p\_cl'_{y}\\
\end{bmatrix}
\end{equation}
\vspace{-1mm}

\begin{algorithm}[h]
\caption{CLAUSE Function}\label{alg:Clause_Compute}
\KwIn{$L$ (Literal), $l\_mask$ (literal mask), $Action$ (TA action, 1 means \textbf{Inc}, 0 means \textbf{Exc}), $cl\_buffer\_mask$ (the clause buffer mask), $x$ (literal width of clause computation matrix), $y$ (clause width of clause computation matrix)}

\SetKwFunction{CLAUSE}{CLAUSE}

\CLAUSE{$TM\_type$, $l\_mask$, $cl\_buffer\_mask$, $h$, $x$, $y$, $L$, $Action$}{
    \ForEach{$i \in y$}{
        \ForEach{$j \in x$}{
            $p\_cl[i] \leftarrow p\_cl[i] \land cl\_buffer\_mask[i] \land ((L[j] \lor l\_mask[j]) \lor \sim Action[i][j])$\;
        }
    }
    \Return{$p\_cl$}\;
}
\end{algorithm}
\setlength{\textfloatsep}{0pt}
\textbf{c) Introducing Partial Clauses:} Instead, drawing from previous works~\cite{Matador, Imbue}, the full clause matrix computation is now decomposed into smaller partial clause matrix computations. For a given datapoint, DTM uses computation slices of a group of clauses that utilize only a subset of all TA actions and their corresponding literals. These are \textit{ partial clauses}. Suppose that a TM model contains $f$ features (so \texttt{$f$$\times$$2$} literals), $c$ clauses per class for Vanilla TM and $c$ clauses altogether for CoTM, and $h$ classes\footnote{These variables will now be used throughout the remainder of the paper and their relationships will be further elaborated in the DTM algorithm blocks. The reader is encouraged to refer back to Fig.~\ref{fig:CoTM_block} and Fig.~\ref{fig:Mem_slice} to better understand the design choices made with the hardware implementation.}: The \texttt{Clause Matrix} will process $x$ of these literals to compute $y$ partial clauses, using \(x \times y\) corresponding TA actions \(x \times y\) for each computation. The next few paragraphs discuss the iterations of this divided computation and how redundancies are handled when the matrices cannot be divided without remainders.

\begin{algorithm}[h]
\caption{CSUM Function}\label{alg:CSUM}
\KwIn{$cl$ (loaded clause), $cl\_mask$ (clause mask in weight computation matrix), $W$ (loaded weights), $m$ (clause width of weight computation matrix), $n$ (class width of weight computation matrix), $L_{csum}$ (Maximum length of class sum)}

\SetKwFunction{CSUM}{CSUM}

\CSUM{$m, n, cl\_mask, cl, W, L_{csum}$}{
    \ForEach{$i \in n$}{
        \ForEach{$j \in m$}{
            $p\_cs[i] \leftarrow p\_cs[i] + (cl[j] \land cl\_mask[j]) \times W[i][j]$\;
        }
    }
    \Return{$p\_cs$}\;
}
\end{algorithm}

\textbf{d) Partial Clause Matrix Computation: }Equation~\ref{eq:clause_compute_matrix} shows how partial clauses are computed in the \texttt{Clause Matrix}. The vector $p\_cl$ represents the result of the previous partial clauses while $p\_cl'$ represents the current partial clauses result. $p\_cl$ will be set to a $1$ vector at the beginning of a group of clause computation. The computation remains the same between the literal $L$ and the TA actions subset $TA^{i}_{j}$ for the $i$th literal in the $j$th clause in this matrix. 

\textbf{e) Partial Clause Iteration Sequence: }Fig.~\ref{fig:mem_clause_slice} shows the iterations for the \texttt{Clause Matrix} block. The matrix needs \(a = \lceil\frac{2 \times f}{x} \rceil\) clock cycles to compute a group of \texttt{y} clauses and \(b = \lceil \frac{c}{y} \rceil\) groups to compute one class of clauses. For CoTM, this is the full iteration as the CoTM will use the shared clause pool (see Fig.~\ref{fig:CoTM_block}), but when computing clauses for Vanilla TM, \(b \times h\) groups of clauses will be computed.

\begin{figure}[htbp]
    \centering
    \begin{subfigure}{0.44\columnwidth}
        \centering
        \includegraphics[width=\linewidth]{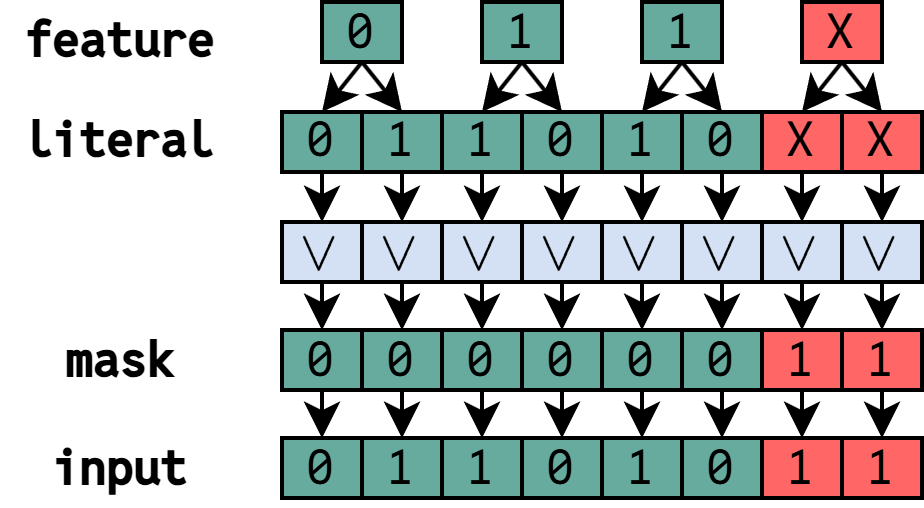}
        \vspace{-5mm}
        \caption{Literal mask (Clause Matrix)}
        \label{fig:literal mask}
    \end{subfigure}
    \begin{subfigure}{0.44\columnwidth}
        \centering
        \includegraphics[width=\linewidth]{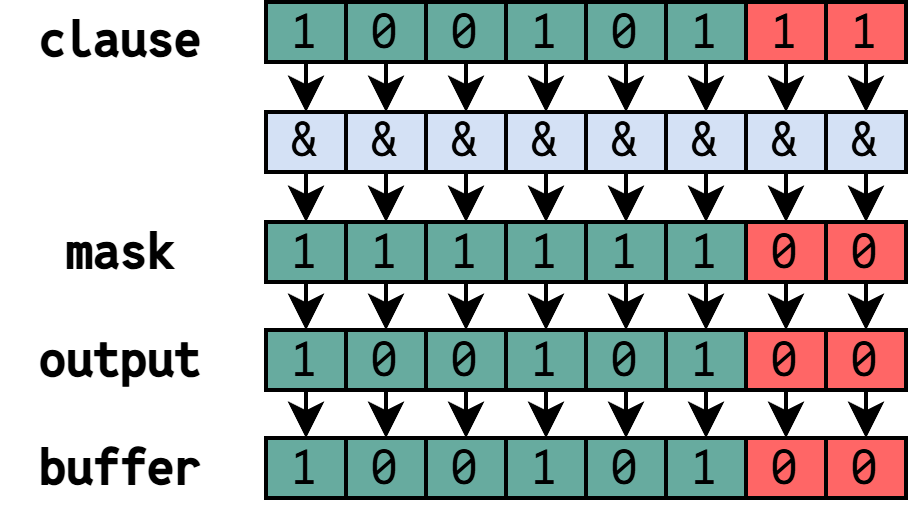}
        \vspace{-5mm}
        \caption{Clause buffer mask}
        \label{fig:clause buffer mask}
    \end{subfigure}


    \begin{subfigure}{0.44\columnwidth}
        \centering
        \includegraphics[width=\linewidth]{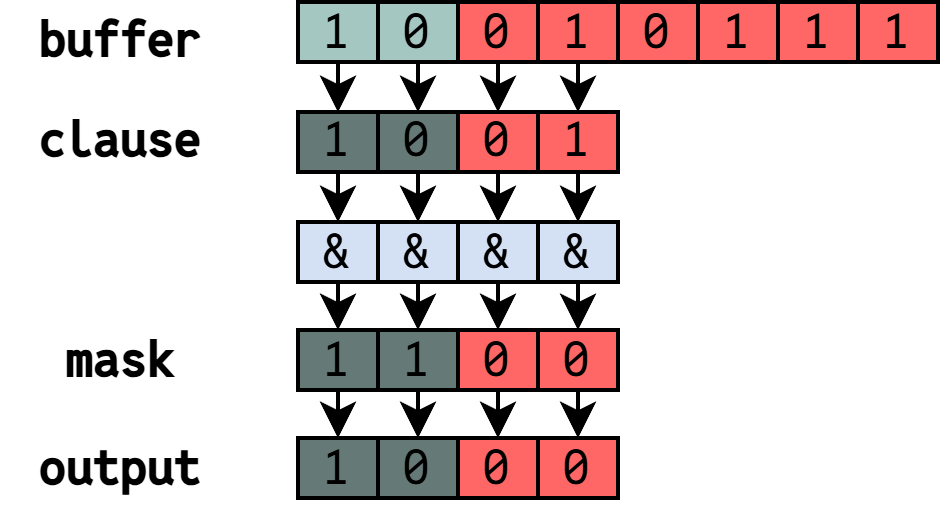}
        \vspace{-5mm}
        \caption{Clause mask (Weight Matrix)}
        \label{fig:clause mask}
    \end{subfigure}
    \begin{subfigure}{0.44\columnwidth}
        \centering
        \includegraphics[width=\linewidth]{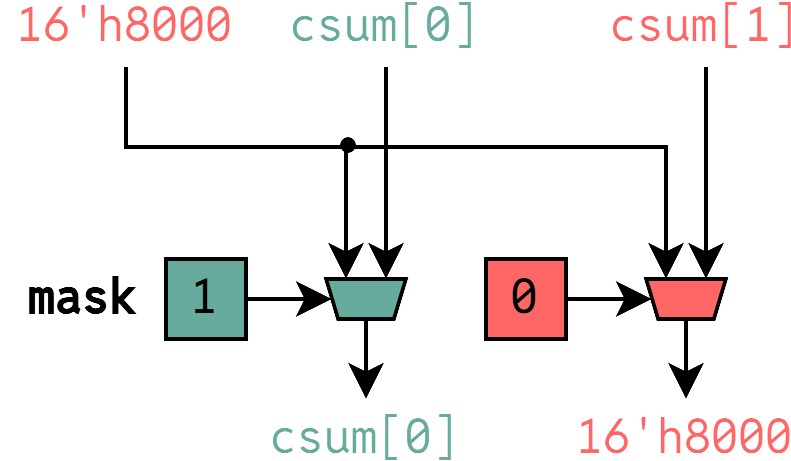}
        \vspace{-5mm}
        \caption{Class mask}
        \label{fig:class mask}
    \end{subfigure}
    \vspace{-1.5mm}
    \caption{\small{Masks for remainder compute elements (marked in red).}}
    \label{fig:subfigures}
\end{figure}

\textbf{f) Masking Remainder Compute: }There may be cases of remainder literals in the last slice fed into the \texttt{Clause Matrix}. In these cases, the remainder literals (\texttt{X}) will be masked. This is shown in Fig.~\ref{fig:literal mask}. An inverted version of this mask is also used for \texttt{TA Update Matrix}. Similarly to the remainder literal mask when there are remainder clauses, a clause mask is used to prevent writing these remainder clauses into the clause buffer (see Fig.~\ref{fig:clause buffer mask}). The mask uses logic~\texttt{0} for these remaining elements so that when logic~\texttt{AND}'d, they will is always~\texttt{0}. All these ideas are culminated in (Algorithm~\ref{alg:Clause_Compute}). 

\textbf{g) Feeding TA actions for inference:} Considering the model size and resource limitations of the target FPGA platform, it is more efficient to store the TA states in BRAM instead of registers. For a clause computation with \(x\) literals and \(y\) clauses, assuming the TA state length is \(L_{TA}\) bits and the BRAM used to store TA states is configured as \(W_{TA\_RAM}\) bits, at least \(\lceil x \times y \times \frac{L_{TA}}{W_{TA\_RAM}}\rceil\) BRAMs will be configured as TA RAMs. The clause computation matrix will read the actions from one row of TA states data in each clock cycle (see Fig.~\ref{fig:CoTM_Arch}-\textbf{1}).  Each clause computation slice in Fig.~\ref{fig:mem_clause_slice} takes one row of on-chip RAM (e.g. BRAM and URAM).

\vspace{-2mm}
\setlength{\textfloatsep}{0pt}
\begin{equation} \label{eq:csum_compute_matrix}
\begin{bmatrix}
    p\_cs_{1}\\
    p\_cs_{2}\\
    \vdots\\
    p\_cs_{n}\\
\end{bmatrix}
+
\begin{bmatrix}
w^{1}_{1} & w^{2}_{1} & \dots & w^{m}_{1}\\
w^{1}_{2} & w^{2}_{2} & \dots & w^{m}_{2}\\
\vdots & \vdots & \ddots & \vdots\\
w^{1}_{n} & w^{2}_{n} & \dots & w^{m}_{n}\\
\end{bmatrix}
\times
\begin{bmatrix}
    cl_{1}\\
    cl_{2}\\
    \vdots\\
    cl_{m}\\
\end{bmatrix}
=
\begin{bmatrix}
    p\_cs'_{1}\\
    p\_cs'_{2}\\
    \vdots\\
    p\_cs'_{n}\\
\end{bmatrix}
\end{equation}

After a group of clauses is computed in the \texttt{Clause Matrix}, the clause outputs will be stored in either registers or BRAM depending on the FPGA resources - this is the clause buffer. The clause buffer stores only \textbf{1} class of clause outputs.

\textbf{h) Class Sum Computation: } A similar slicing strategy is also applied in the \texttt{Weight Matrix} for class sum computation (see Fig.~\ref {fig:CoTM_Arch}). Loading all weights and clauses to compute one class sum is resource-intensive. 

\textbf{i) Partial Class Sums:} Splitting class sums into multiple partial matrix multiplications provides better performance (running at higher frequencies) and lower resource costs. Thus, full class sum computation is split into multiple rounds of partial weight multiplication and vector addition. The weight matrix computes \(n\) partial class sums using the \(m\) clauses in each iteration (see Equation~\ref{eq:csum_compute_matrix}).

In Equation~\ref{eq:csum_compute_matrix}, $p\_cs$ represents the previous partial class sum result, $p\_cs'$ represents the current partial class sum result.$p\_cs$ will be set to $0$ when start a group of class sum computation, $cl_{j}$ represents the $p$th clause in this class, and $w^{j}_{i}$ represents the weight of the $j$th clause of the $i$th class in this matrix.

For the CoTM, the accelerator takes \(p = \lceil \frac{c}{m} \rceil\) clock cycles to compute a group of \(n\) class sums, and \(q = \lceil \frac{h}{n} \rceil\) groups of partial class sums to get all class sums for $h$ classes.

For Vanilla TM class sum computation, the weights become~\textbf{+1} for positive clauses and~\textbf{-1} for negative clauses. Positive clause are even indexed while negative clauses odd indexed. Therefore, a class sum for class $i$ can be obtained by Equation~\ref{eq:vanilla_csum_compute_matrix} using the same dimension vectors and matrix as in Equation~\ref{eq:csum_compute_matrix}, where $p\_cs_{i}$ and $p\_cs'_{i}$ represent the current and the next partial class sum of the $i$th class, and $cl^{i}_{j}$ represent the $j$th clause in the $i$th class. In this case, the weights loaded into Fig.~\ref{fig:CoTM_Arch}-~\textbf{2} become $1$ and $-1$ for even clauses and odd clauses respectively. The accelerator takes \(p = \lceil \frac{c}{m} \rceil\) clock cycles to compute one full class sum for the Vanilla TM.
\vspace{-1mm}
\begin{equation} \label{eq:vanilla_csum_compute_matrix}
\begin{bmatrix}
    p\_cs_{i}\\
    p\_cs_{i}\\
    \vdots\\
    p\_cs_{i}\\
\end{bmatrix}
+
\begin{bmatrix}
1 & -1 & \dots & -1\\
1 & -1 & \dots & -1\\
\vdots & \vdots & \ddots & \vdots\\
1 & -1 & \dots & -1\\
\end{bmatrix}
\times
\begin{bmatrix}
    cl^{i}_{1}\\
    cl^{i}_{2}\\
    \vdots\\
    cl^{i}_{m}\\
\end{bmatrix}
=
\begin{bmatrix}
    p\_cs'_{i}\\
    p\_cs'_{i}\\
    \vdots\\
    p\_cs'_{i}\\
\end{bmatrix}
\end{equation}

\textbf{j) Feeding the Weights: }Considering the size of all weights, it is more feasible to store them in BRAM on FPGA. For an accelerator with a weight matrix sized as $m \times n$, assume that each weight has \texttt{w} bits, and that the BRAM has used to store the weights are configured as \texttt{u} bits. Then \(\lceil \frac{m \times n \times w}{u} \rceil\) BRAMs are required for weight storage. When computing a group of class sums, only one row of data will be read as the weights for the computation in each clock cycle (see  Fig.~\ref{fig:CoTM_Arch}-\textbf{2}). Vanilla TM inference only requires constant weights, so there is no reading from BRAM.

\textbf{k) Masking Remainder Compute: }When there is a remainder clause in \texttt{Weight Matrix}, a clause mask is applied with logic~\texttt{0} for the remainder bits and logic~\textbf{AND}'d with the remainder clause bits so they will not contribute to the class sum (see Fig.~\ref{fig:clause mask}). When there are remainder class sums computed, a class sum mask is applied as post-processing which assigns $-2^{L_{csum}-1}$ in binary ($L_{csum}$ is the maximum length of the class sum in binary, and $-2^{L_{csum}-1}$ is the minimum possible class sum in binary) to the remainder class sums while the others remain, so that the remainder class sum will never be greater than used classes (see Fig.~\ref{fig:class mask}).

\textbf{l) Class Sum and Argmax: }When computing the class sums for CoTM, the binary multiplication is optimized to a logic~\texttt{AND} operation between the clause and the corresponding weight bit by bit (see Algorithm~\ref{alg:CSUM}). The \texttt{Argmax} is a comparison tree with $n$ class sums as input. The maximum class sum and its class index will be cached and then used for comparison with the maximum class sum in the next group of class sums until all class sums are compared. 

\textbf{\textit{B. DTM Training Modules:}} This subsection presents the implementation details of the training modules for the DTM. It is recommended to read this section and its algorithms with reference to Fig.~\ref{fig:TM_feedback}\footnote{For example if a paragraph has a bold label Class-level feedback - view this part of Fig~\ref{fig:TM_feedback} against the same algorithm name in this section. This should make it easier to follow the nested condition parts in the algorithms}. Much like in the background section, this subsection presents how the feedback to TAs is passed from class level to clause level to TA level.    

\begin{algorithm}[h]
\caption{Class-level Feedback Generation}
\label{alg:NCC}
\KwIn{$T$ (hyper parameter), $y_c$ (input class, $1$ means target class, $0$ means negated class), 
$c\_rand$ (random number for negated class choice), $L_{w\_rand}$ (random number length for clause-level feedback compute), $Target\_Class$ (target class index), $Update\_csum$ (Class sum for updated class), $Class\_num$ (Class number)}

\SetKwFunction{NCGen}{NC\_Gen}
\SetKwFunction{UpdateProbabilityCompute}{Update\_Probability\_Compute}

\NCGen{$c\_rand$, $Target\_Class$, $Class\_num$}{
    $RN_C = c\_rand \% (Class\_num - 2)$\;
    \eIf{$RN_C < Target\_Class$}{
        $Negated\_Class = RN_C$\;
    }{
        $Negated\_Class = RN_C + 1$\;
    }
    \Return{$Negated\_Class$}\;
}

\BlankLine
\UpdateProbabilityCompute{$y_c$, $c\_rand$, $L_{w\_rand}$, $Target\_Class, Class\_num$, $Update\_csum$}{
    \BlankLine
    \eIf{$y_c == 1$}{
        $Update\_csum = cum[Target\_Class]$\;
    }{
        $Negated\_index$ = \NCGen{$c\_rand$, $Target\_Class$, $Class\_num$}\;
        $Update\_csum = cum[Negated\_index]$\;
    }
    $CSum = clip(Update\_csum, [-T, T])$\;
    \eIf{$y_c == 1$}{
        $P_{Cl\_Update} = (T - CSum) \times 2^{L_T - 1}$\;
    }{
        $P_{Cl\_Update} = (T + CSum) \times 2^{L_T - 1}$\;
    }
    \Return{$P_{Cl\_Update}$}\;
}

\end{algorithm}

\textbf{a) Class-Level Feedback: } Training requires both the target class and a randomly chosen class (referred to as the negated class) for each training datapoint. The target class is padded into the input data stream sent to the accelerator (presented later), while the randomly chosen negated class is determined within the hardware accelerator (see Function~\textbf{NC\_Gen} in Algorithm~\ref{alg:NCC} (\texttt{NC\_Gen})). The clause-level update probability is computed in Function~\textbf{UpdateProbabilityCompute} with some changes to be more easily translated to hardware.

\begin{algorithm}[h]
\caption{Clause Level Feedback Generation}
\label{alg:clause_update}
\KwIn{$TM\_type$ (TM type, $2'b01$ represents Vanilla TM and $2'b10$ represents CoTM), $index$ (index of updated class), $P_{Cl\_Update}$ (Update\_probability), $T$ (hyper parameter), $cl$ (Clause), $cl\_mask$ (clause mask for weight computation matrix), $m$ (clause number in weight computation matrix), $W$ (Signed weight of this clause), $w\_rand$ (random number for clause-level feedback compute), $y_c$ (the updated class, $1$ represents target class, $0$ represents negated class)}
\SetKwFunction{ClauseUpdate}{Clause\_Update}
\ClauseUpdate{$cl$, $W$, $cl\_mask$, $w\_rand$, $y_c$}{
    \ForEach{$i \in m$}{
        \If{$cl\_mask[i] == 1$}{
            \If{$P_{Cl\_Update} \times T \geq w\_rand$}{
                \If{$y_c == 1$}{
                    \If{$W[index][i] < 0$}{
                        $Feedback[i] = 2'b01$\;
                    }
                    \ElseIf{$W[index][i] \geq 0$}{
                        $Feedback[i] = 2'b10$\;
                    }
                    \If{$cl[i] == 1$}{
                        \If{$TM\_type == 2'b10$}{
                            $W[index][i] = W[index][i] + 1$\;
                        }
                    }
                }
                \If{$y_c == 0$}{
                    \If{$W[index][i] \geq 0$}{
                        $Feedback = 2'b01$\;
                    }
                    \ElseIf{$W[index][i] < 0$}{
                        $Feedback = 2'b10$\;
                    }
                    \If{$cl[i] == 1$}{
                        \If{$TM\_type == 2'b10$}{
                            $W[index][i] = W[index][i] - 1$\;
                        }
                    }
                }
            }
        }
    }
    \Return{$Feedback, W$}\;
}
\end{algorithm}

\textbf{b) Clause-Level Feedback: } After selecting a class to be updated ($y_c$), the clause-level feedback should be determined based on the comparison result between the product of a \(L_{w\_rand}\)-bit random number \(w\_rand\) with the hyperparameter $T$ and the clause update probability (see Algorithm~\ref{alg:NCC}). 

The clause-level feedback will be stored in \texttt{Clause Feedback Buffer} as either registers or BRAM. In this architecture, for better simplicity of design, the size of the \texttt{Weight Update Matrix} (see Fig.~\ref{fig:CoTM_Arch}-\textbf{3}) matches the \texttt{Weight Matrix}. Thus, in each clock cycle, clause-level feedbacks for \(m\)  clauses will be computed, and \(m\) weights for the corresponding clauses will be updated. The weights update only in CoTM mode. Algorithm~\ref{alg:clause_update} shows the clause-level feedback and weight updates within \texttt{Weight Update Matrix} in~\ref{fig:CoTM_Arch}-~\textbf{3}. The clause mask from Section~\ref{subsec:inference} is reused to prevent remainder clauses from updating.

\begin{algorithm}[h]
\caption{Group Clause TA Update}
\label{alg:group_TA_update}
\KwIn{$P_{TA\_Update}$ (TA update probability), $Feedback$ (loaded Clause level feedback), $L$ (loaded literal), $l\_mask$ (literal mask), $TA$ (loaded unsigned TA state), $action$ (TA action of the loaded TAs), $L_{TA\_rand}$ (random number length), $TA\_rand$ (random number), $cl$ (loaded clause), $cl\_buffer\_mask$ (the clause buffer mask), $x$ (literal width of clause computation matrix), $y$ (clause width of clause computation matrix)}
\SetKwFunction{TAFD}{TAFD}

\TAFD{\makebox[0pt][l]{$x, y, L, TA, cl, TA\_rand, P_{TA\_Update}, Feedback,$} \newline $l\_mask, cl\_buffer\_mask, action$}{

    \ForEach{$i \in y$}{
        \ForEach{$j \in x$}{
            \If{$cl\_buffer\_mask[i] \land l\_mask[j]$}{
                \If{$Feedback[i] == 2'b01$}{
                    \If{$\sim cl[i] \lor (cl[i] \land \sim L[j])$}{
                        \If{$P_{TA\_Update} \geq TA\_rand$}{
                            \If{$TA[i][j] > 0$}{
                                $TA[i][j] = TA[i][j] - 1$\;
                            }
                        }
                    }
                    \ElseIf{$cl[i] \land L[j]$}{
                        \If{$P_{TA\_Update} < TA\_rand$}{
                            \If{$TA[i][j] < 2^{L_{TA}} - 1$}{
                                $TA[i][j] = TA[i][j] + 1$\;
                            }
                        }
                    }
                }
                \ElseIf{$Feedback[i] == 2'b10$}{
                    \If{$cl[i] \land \sim L[j] \land action[i][j] == 1$}{
                        \If{$TA[i][j] < 2^{L_{TA} - 1}$}{
                            $TA[i][j] = TA[i][j] + 1$\;
                        }
                    }
                }
            }
        }
    }
    \Return{$TA$}\;
}
\end{algorithm}
\setlength{\textfloatsep}{0pt}
\begin{algorithm}[h]
\caption{Optimized TA Update}\label{alg:TA_update_opt}
\KwIn{$Feedback$ (loaded feedback for the current group of clauses)}

\SetKwFunction{TAFDOPT}{TAFD\_OPT}

\TAFDOPT{$x, y, L, TA, action, cl, TA\_rand, Feedback,$ \newline
        $P_{TA\_Update}, l\_mask, cl\_buffer\_mask$}{
    
    \While{TA update not finished}{
        \If{$\exists i \in y$ such that $Feedback[i] \neq 2'b00$}{
            \SetKwFunction{TAFD}{TAFD}
            \TAFD{$x, y, TA, action, TA\_rand, P_{TA\_Update}, L,$ \newline 
                  $Feedback, cl, l\_mask, cl\_buffer\_mask$}\;
        }
        \Else{
            Allocate TA RAM address to the start of the next group of clauses\;
        }
    }
    \Return{$TA$}\;
}
\end{algorithm}

\textbf{c) TA-Level Feedback: }Similarly, to match the incoming data rate when computing the clauses, the \texttt{TA Update Matrix} is designed to be the same size as the \texttt{Clause Matrix}. Hence, in each clock cycle, \(y\) clause-level feedback will be read from the buffer registers. Then \(y\) corresponding clauses, \(x\) features, and \(x \times y\) \(L_{TA\_rand}\)-bit random numbers will be fed into the \texttt{TA Update Matrix} where \(TA\_rand\) is the random number for the TA-level feedback computation, and \(L_{TA\_rand}\) is the length of these random numbers. In the accelerator, \(s\) is defined as \(L_s\) bits. To reduce resource utilization, the TA update probability \(\frac{1}{s}\) is precomputed, because the random numbers generated in the hardware are \(L_{TA\_rand}\)-bit binary numbers so that the TA update probability $P_{TA\_Update}$ becomes \(\frac{2^{L_{TA\_rand}}}{s}\) (see Algorithm~\ref{alg:group_TA_update}). Fig.~\ref{fig:CoTM_Arch}-\textbf{4} shows the architecture of the \texttt{TA Update Matrix}, it contains \(x \times y\) TA update blocks, and for the TM update, the TA update block takes $a$ clock cycles to update the TAs in a group of $x$ clauses, and TAs in one class require $a \times b$ clock cycles. The inverted literal mask (Section~\ref{subsubsec:feature buffer}) and clause buffer mask (Section~\ref{subsec:Clause_matrix}) become control signals preventing updates of remainder TAs in a clause computation slice. Function~\textbf{TAFD} in Algorithm~\ref{alg:group_TA_update} is used to update the TAs in one TA slice.

\begin{figure}[h]
    \centering
    \includegraphics[width= 0.98\linewidth]{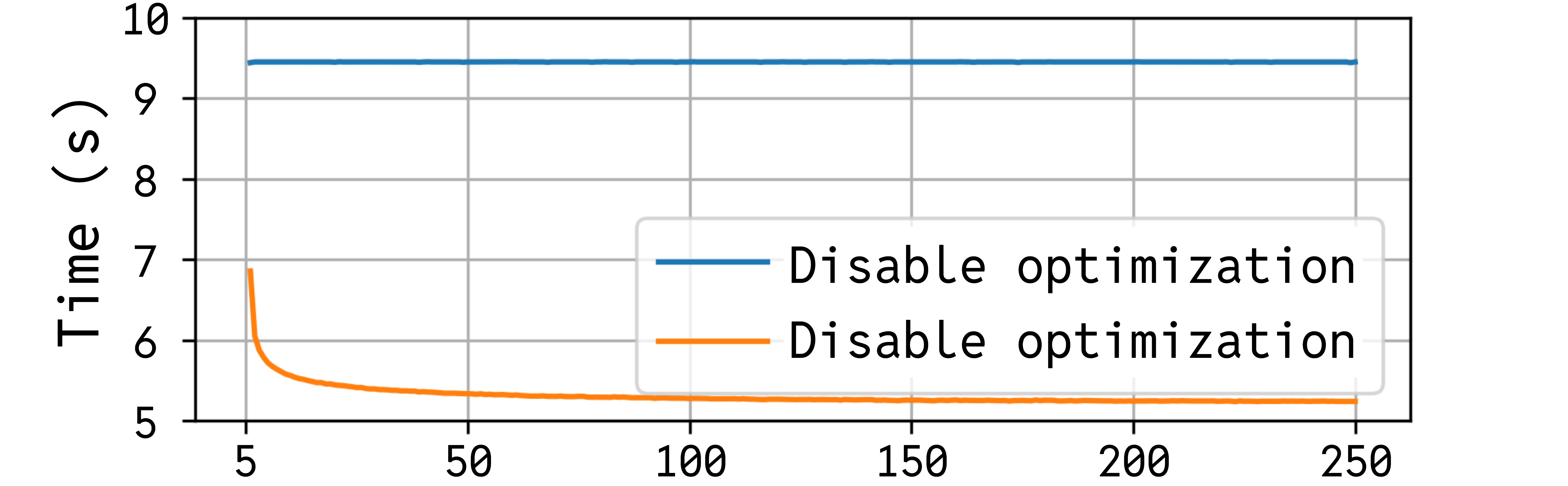}
    \caption{\small{Training time per epoch.}}
    \label{fig:MNIST_opt_time}
\end{figure}

\textbf{d) Clause-Level Feedback Optimization:} In the proposed architecture, the TAs in a group of clauses are updated concurrently. As the model converges while training, the frequency of feedback to clauses reduces~\cite{REDRESS}. DTM skips loading of all TA slices from BRAM and updation when there is no clause-level feedback to a particular group of clauses, saving precious cycles. Explanation: when the clause matrix slides through the slices (see Fig.~\ref{fig:mem_clause_slice}), if no feedback is present for the corresponding $y$ clauses, the computation slice moves directly to the next row without traversing the current row's bottom column. This is implemented by adjusting the read/write addresses of the TA RAM, as shown in Algorithm~\ref{alg:TA_update_opt}. With this optimization, the CoTM gets $\approx 40\%$ improvement in training time as shown in Fig.~\ref{fig:MNIST_opt_time}, while achieving similar test accuracy. The reduction in training time indicates fewer clause updates each epoch. The training time eventually saturates to a constant value, which approximately equals the inference time plus the weight update time for both target and negated class updating. 
\setlength{\textfloatsep}{0pt}

\begin{figure}[t]
    \vspace{-6mm}
    \centering
    \includegraphics[width = 0.85\linewidth]{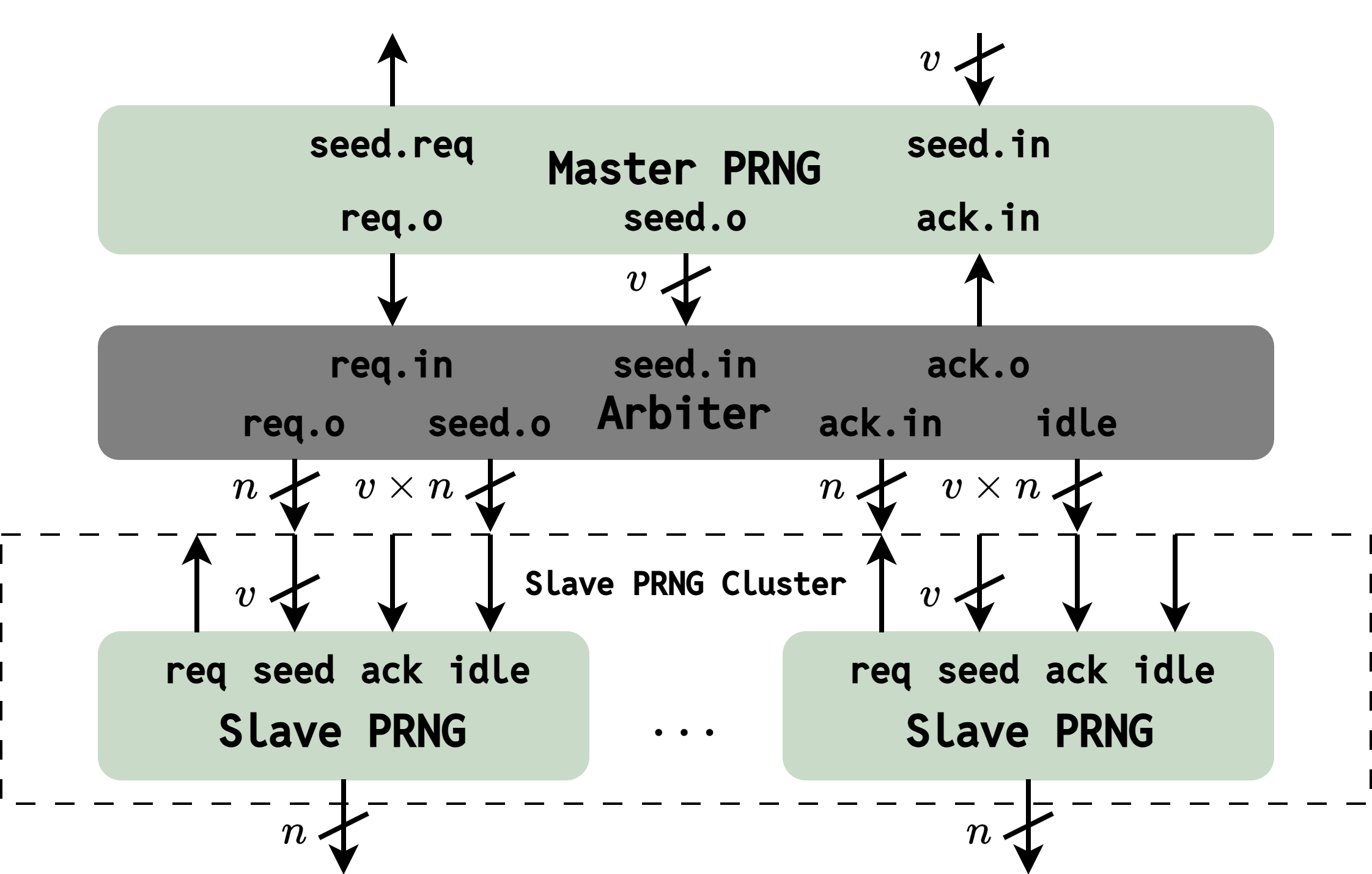}
    \vspace{-1.5mm}
    \caption{\small{The block diagram of the PRNG cluster.}}
    \label{fig:CoTM_PRNG}
\end{figure}

\textbf{\textit{C. Pseudo Random Number Generator (PRNG):}} To generate a large number of random numbers for parameter updating in real time, the accelerator is integrated with a master-slave architecture PRNG cluster developed from~\cite{10455073}. In the PRNG cluster, the master PRNG is used to generate the seeds for the slave PRNGs. For each slave PRNG, if the random number generation finishes one PRNG cycle period, it will request a new seed from the master PRNG; this is called seed refresh. Fig.~\ref{fig:CoTM_PRNG} shows the architecture of the PRNG cluster. The authors in~\cite{10455073} map the PRNG cluster to DSP blocks; however, the slave PRNG used in this design is LFSR based due to DSP block limitations in the chosen FPGAs. 

In the LFSR version of the master-slave PRNG cluster, the master PRNG will generate and set the seeds for the slave PRNG after $2^{L_{LFSR}}$ clock cycles, where $L_{LFSR}$ is the length of LFSR. One DSP slice can only afford $16$-bits Mid-square based PRNG, hence in this accelerator, LFSR is used as the slave PRNG in the PRNG cluster. The processor supplies and programs the master PRNG seed in real time.

\begin{figure}[h]
    \centering
    \includegraphics[width =0.9\linewidth]{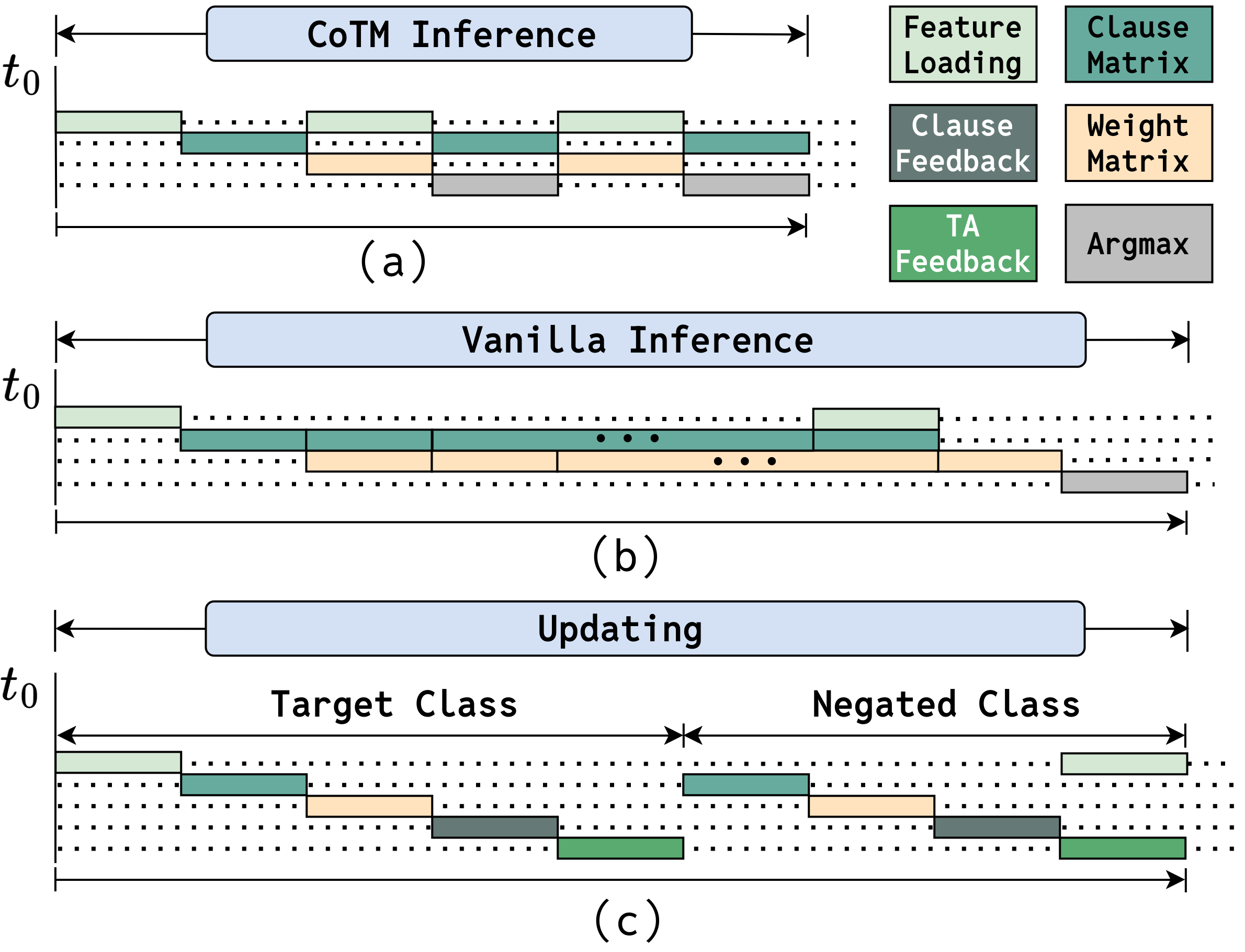}
    \vspace{-2.5mm}
    \caption{\small{Accelerator Timing: a) timing diagram for CoTM inference; b) timing diagram for Vanilla TM inference; c) timing diagram for DTM training (same procedure for both Vanilla and CoTM).}}
    \label{fig:TM_inference_timing}
\end{figure}

\begin{figure*}[h]
    \centering
    \begin{subfigure}[b]{0.53\textwidth}
        \includegraphics[width = \textwidth]{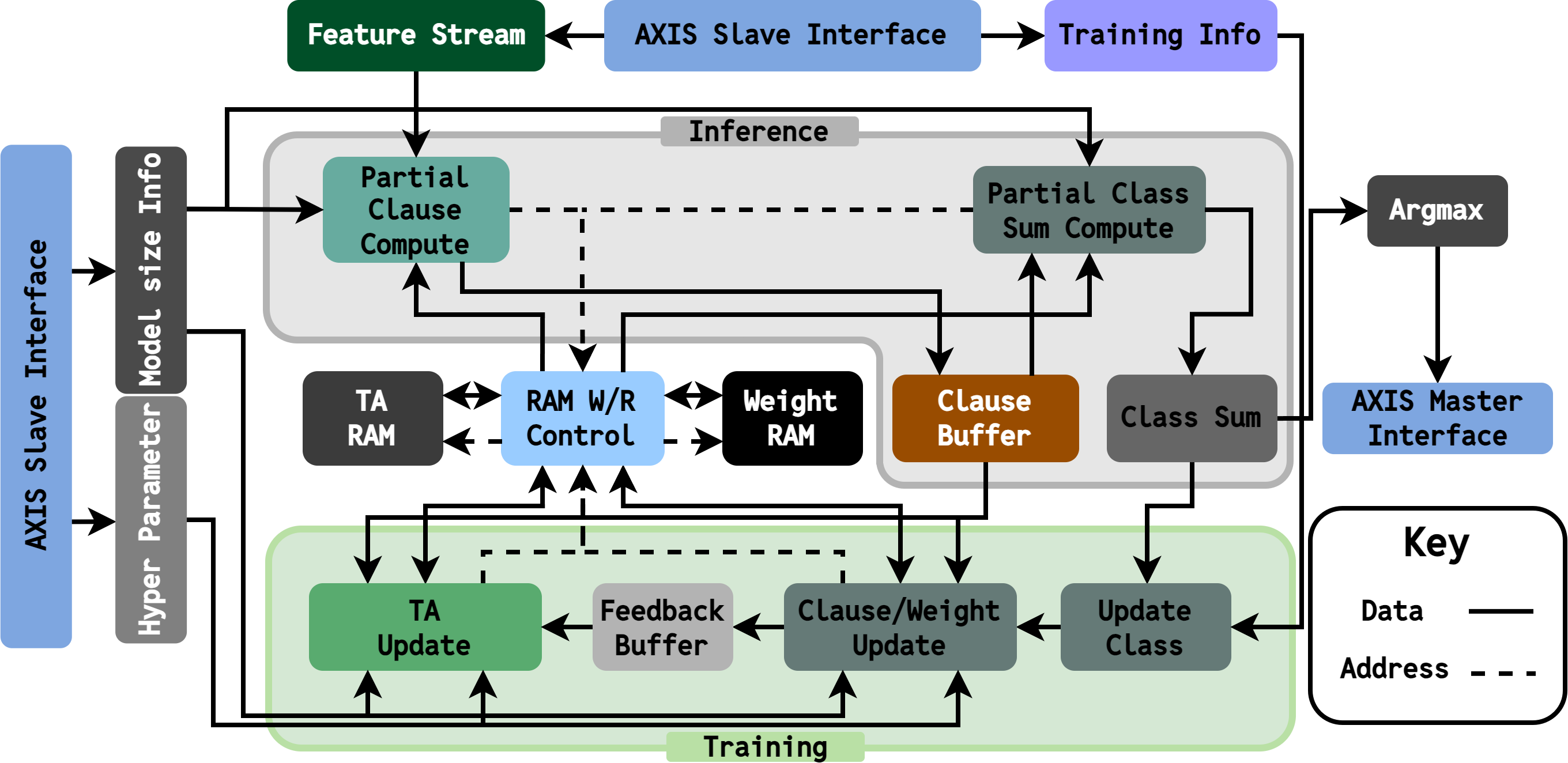}
        \caption{The data flow in IP core.}
        \label{fig:IP_data_flow}
    \end{subfigure}
    \hfill
    \begin{subfigure}[b]{0.39\textwidth}
        \includegraphics[width = 0.9\textwidth]{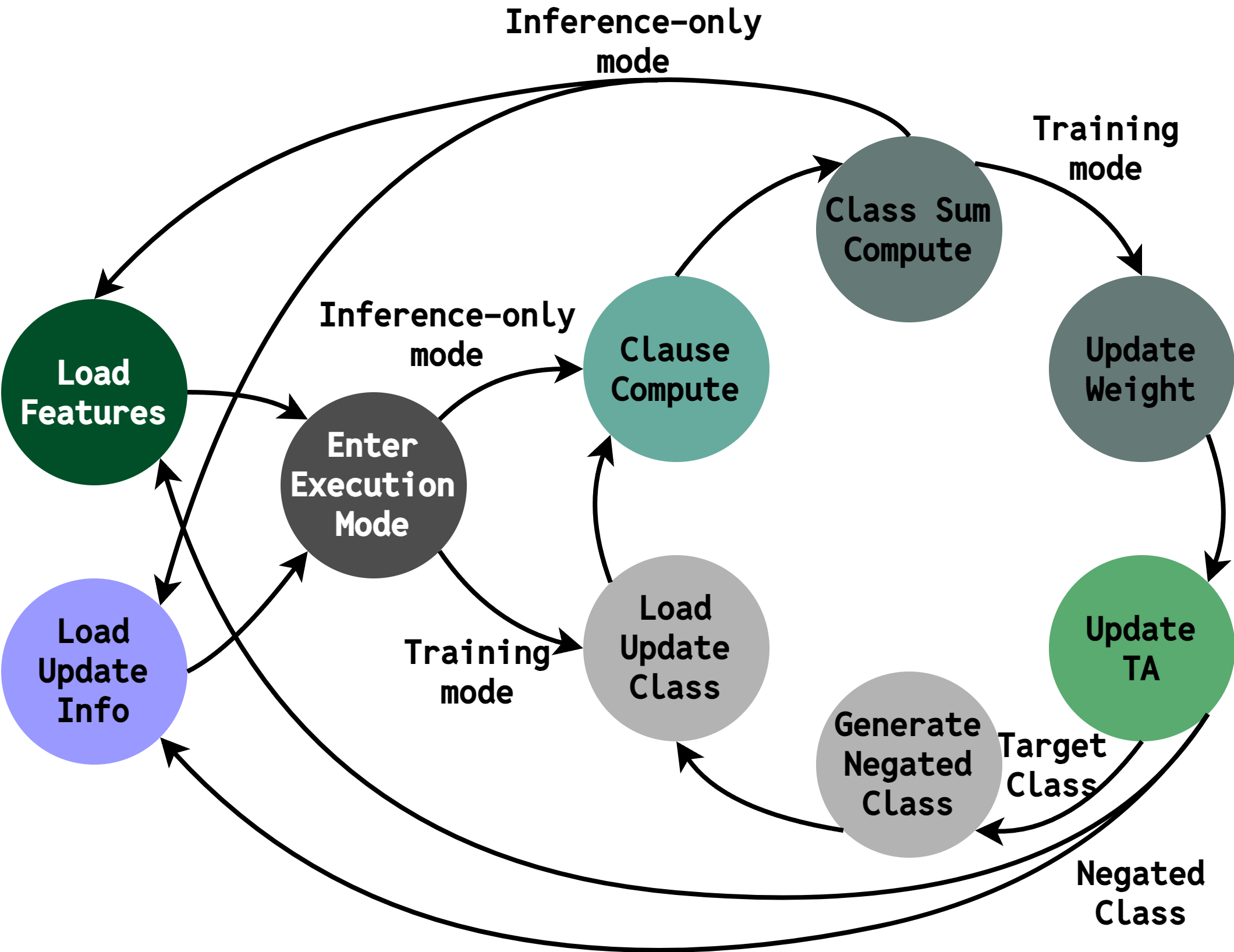}
        \caption{The process flow of the proposed architecture.}
        \label{fig:CoTM_Ctrl_Flow}
    \end{subfigure}
    \vspace{-1mm}
    \caption{\small{DTM accelerator data flows.}} 
    \label{fig:CoTM_Data_Flow}
\end{figure*}

\textit{\textbf{D. DTM Control and Data Flows:}} \textbf{a) Programming:} Before inference or training, the configuration data including the TM type, feature number, clause number and class number are sent to the accelerator, allowing it to calculate clause and weight compute rounds. Upon receiving these data, the necessary clauses and weight masks are derived, as outlined in~\ref{subsec:inference}.For training, hyperparameters like the TA state update probability \(\frac{1}{s}\) and threshold $t$ are provided. The accelerator calculates \(\frac{1}{s}\), stores $t$, and initializes the TA states and weights in RAM using PRNGs once the model size is configured. The programming data (model size, PRNG seeds, and hyperparameters) are sent via an AXI-stream channel. The execution mode (inference or training) and the target class index are embedded within the feature data stream. Fig.~\ref{fig:IP_data_flow} shows the AXIS integration and how the instruction and feature data flow through the system for the inference and training processes.

\textbf{b) Class Sum Computation:} During training, only the target class sum is computed; in inference, class sums for all classes are calculated. The feature data are stored when received; the system uses this to identify the mode (training/inference), and loads the target class if in training mode. The TA RAM controller allocates the start clause slice, and the \texttt{Clause Matrix} loads TA actions and literals to compute clauses in block~\textbf{1}. Inference is pipelined: When one class of clause computation completes and the weight computation matrix is idle, the weight RAM controller allocates the weight slice, loading weights, and clauses to compute class sum in block~\textbf{2}. In inference-only mode, once classes are processed, a new data point is handled. The timing diagrams for the Vanilla TM and CoTM inference are shown in Figs.~\ref{fig:TM_inference_timing}-\textbf{a} and~\ref{fig:TM_inference_timing}-\textbf{b}. 



\textbf{c) Parameter Update:} In training, updates start with class-level feedback from the Clause Update Control (block~\textbf{5} in Fig.~\ref{fig:CoTM_Arch}), which generates clause-level update probabilities. The Training Accelerator updates weights and generates clause feedback (block~\textbf{3}). For CoTM, the weight RAM controller loads the class slice, updates weights, and writes them back to RAM. TA states update in block~\textbf{4}, where the TA RAM controller allocates the clause slice, feeding TAs and literals to the TA update matrix. Updated TA states are written to RAM. TM updates proceed in two rounds: first, the target class, then a random class. Fig.~\ref{fig:TM_inference_timing}-\textbf{c} shows this update flow. After updating, inference resumes by updating class sums, weights, and TA states, while the controller, stream channel, and buffer reset for the next feature stream. The state graph of the system during the training and inference process is shown in Fig.~\ref{fig:CoTM_Ctrl_Flow}. 


\section{Evaluation}
\label{sec:result}
 The DTM accelerator is designed to train on edge applications, particularly cases where recalibration is required. This is typical with IoT sensor data inputs for edge tasks. 
 \textbf{Three Evaluation Points:} This section presents three different evaluation points: firstly, comparing the accelerator against the state of the art in terms of the ability to accelerate the operations per second within a given power budget and associated energy and resource overheads; secondly, the scalability of the design to \textit{both} larger and smaller FPGA platforms when targeting the kind of IoT edge datasets this design should handle; and finally, discussing the design trade-offs when adjusting weight precision and the LFSR length in the PRNG design. The design mapped on PYNQ-Z1 with Xilinx XC7Z020 SoC is referred to as  DTM-S while the design mapped on ZCU-104 with Xilinx ZU-7EV SoC is referred to as DTM-L. Two designs are implemented through AMD's (Xilinx) Vivado Design Suite. This section reports accuracy from 250 training epochs, latency from average training time per data point, and the power from Vivado's implementation reports. The resource utilization of each block and latency in each stage for the DTM-L configuration is presented in Fig.~\ref{fig:Utilization} and Fig.~\ref{fig:Latency}.
 
\textbf{Dataset Rationale: }MNIST~\cite{6296535}, FMNIST~\cite{xiao2017fashionmnistnovelimagedataset}, KMNIST~\cite{clanuwat2018deep}, Google Speech Commands Keyword Spotting with 6 Keywords (yes, no,
up, down, left, right - referred to as KWS-6)~\cite{warden2018speech} (Booleanized as per~\cite{jlpea11020018}) are used as evaluation datasets. Often, deployed models need re-calibration to data affected by sensor aging, temperature, humidity, and other environmental changes~\cite{Concept_Drift}. Therefore, the ability to re-train and adapt, \textit{in situ}, is more energy-efficient than continuous offsite cloud transfers. Additionally, personalized edge training keeps user data private and secure. The MNIST datasets offer benchmarking insights against other designs (Table.~\ref{tab:comparison}, including both system and IP power of DTM), while the KWS6 dataset represents an edge IoT sensor task requiring in-field recalibration and personalized training - this is the type of application this accelerator is designed for (Table.~\ref{tab:date_set}).

\begin{figure}[t]
    \centering
    \includegraphics[width =0.94\linewidth]{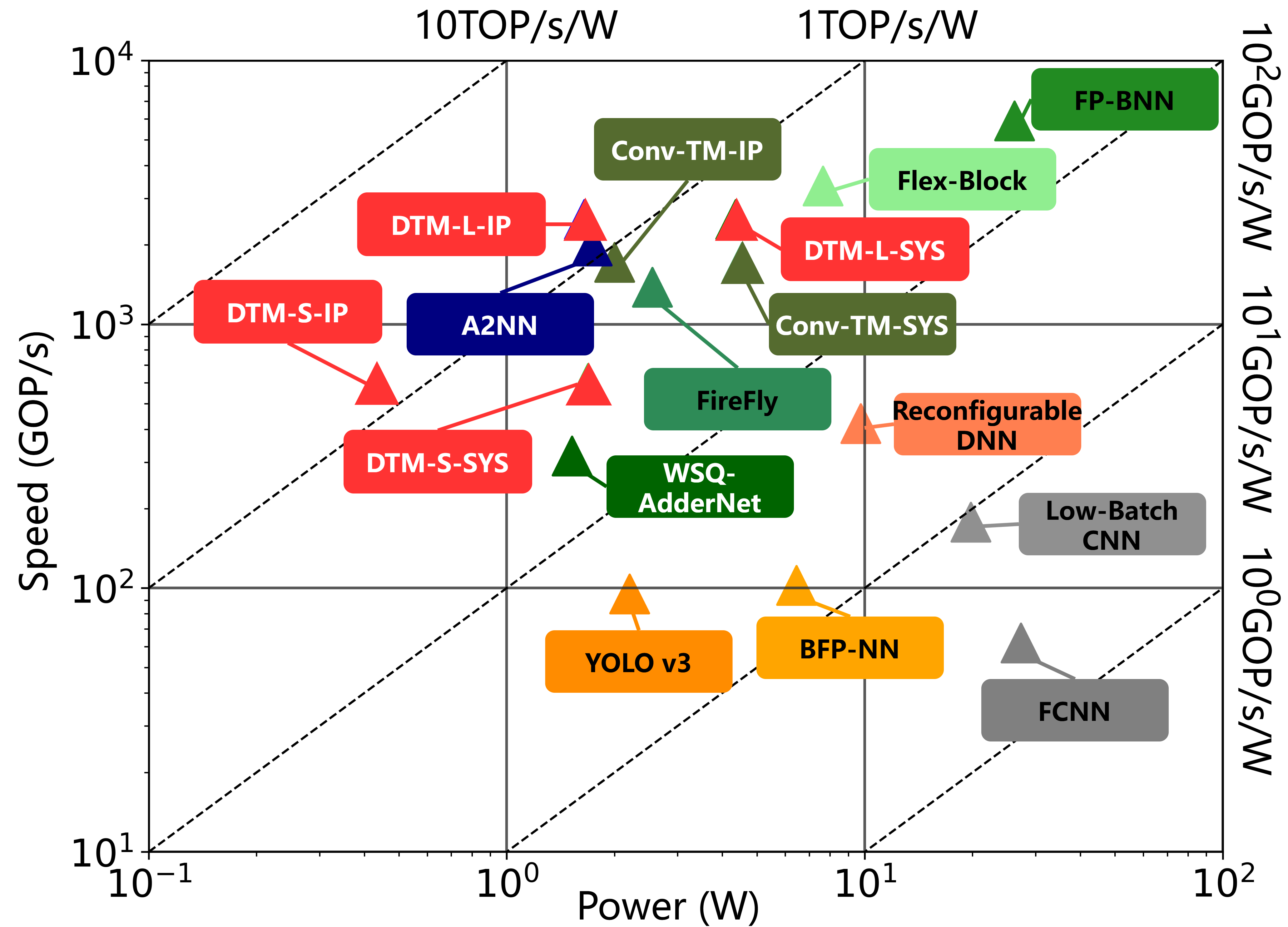}
    \vspace{-1.5mm}
    \caption{\small{Closest comparable State of The Art designs that implement ML \textit{training} on FPGAs and report GOP/s. DTMs shown in red.}}
    \label{fig:GOPs}
\end{figure}

\textbf{Point 1: Comparison with State-of-the-Art: }The DTM designs with different matrix sizes and platforms are compared in terms of Giga-Operations per second per Watt (GOP/s per W) against the related works described in Section~\ref{sec:related_works} through Fig.~\ref{fig:GOPs}: Conv-TM accelerator~\cite{Svein_Conv}, FP-BNN~\cite{FP_BNN}, Reconfig DNN~\cite{reconfigurable_dnn}, Low-batch CNN~\cite{9256704}, F-CNN~\cite{F_CNN}, YOLO v3~\cite{YOLOv3}, BFP-NN~\cite{BFP-NN}, FireFly~\cite{FireFly}, A2NN~\cite{A2NN}, WSQ-AdderNet~\cite{WSQ}, and Flex Block~\cite{FlexBlock}. For DTM and Conv TM design, system-level power including ARM core is reported, named as DTM-SYS and Conv TM-SYS respectively while IP-only power is shown as DTM-IP and Conv TM-IP.

The parameterized matrix sizes in the DTM design allow for scalability as seen through {DTM-S and DTM-L (Table I).} The DTM design philosophy is to utilize larger matrix sizes on larger FPGAs. Therefore, the clause compute matrix size and weight compute matrix size for these two designs are: DTM S - \(32\) literals \(\times\) \(16\) clauses, \(2\) clauses \(\times\) \(4\) classes and DTM L - \(32\) literals \(\times\) \(27\) clauses, \(8\) clauses \(\times\) \(4\) classes. The LFSR length for these two designs is \(12\)-bits and \(24\)-bits, respectively. The training speed, however, depends on TM model sizes and frequency. The accelerators run at 50 MHz and 100 MHz, respectively, due to the FPGAs being based on different technology nodes. Using GOP/s in Fig.~\ref{fig:GOPs} allows for gauging the memory and compute bounds of different implementations. DTM designs are second only to FP-BNN (inference only) and Flex Block in GOP/s. YOLO v3~\cite{YOLOv3}, and BFP-NN~\cite{BFP-NN} have fewer GOP/s compared to the DTMs but also target more complex datasets.

\begin{figure}[t]
    \centering
    \includegraphics[width = \linewidth]{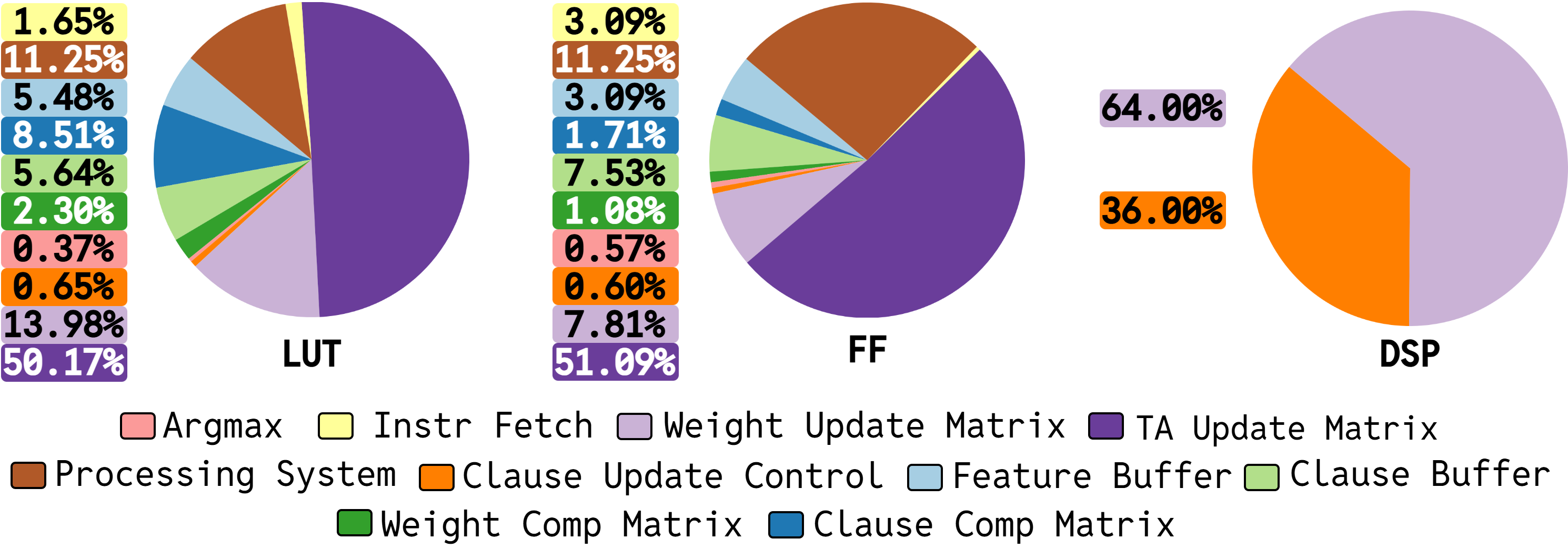}
    \vspace{-6mm}
    \caption{\small{Resource utilization of each module in DTM-L.}}
    \label{fig:Utilization}
\end{figure}

\begin{figure}[t]
    \centering
    \includegraphics[width = \linewidth]{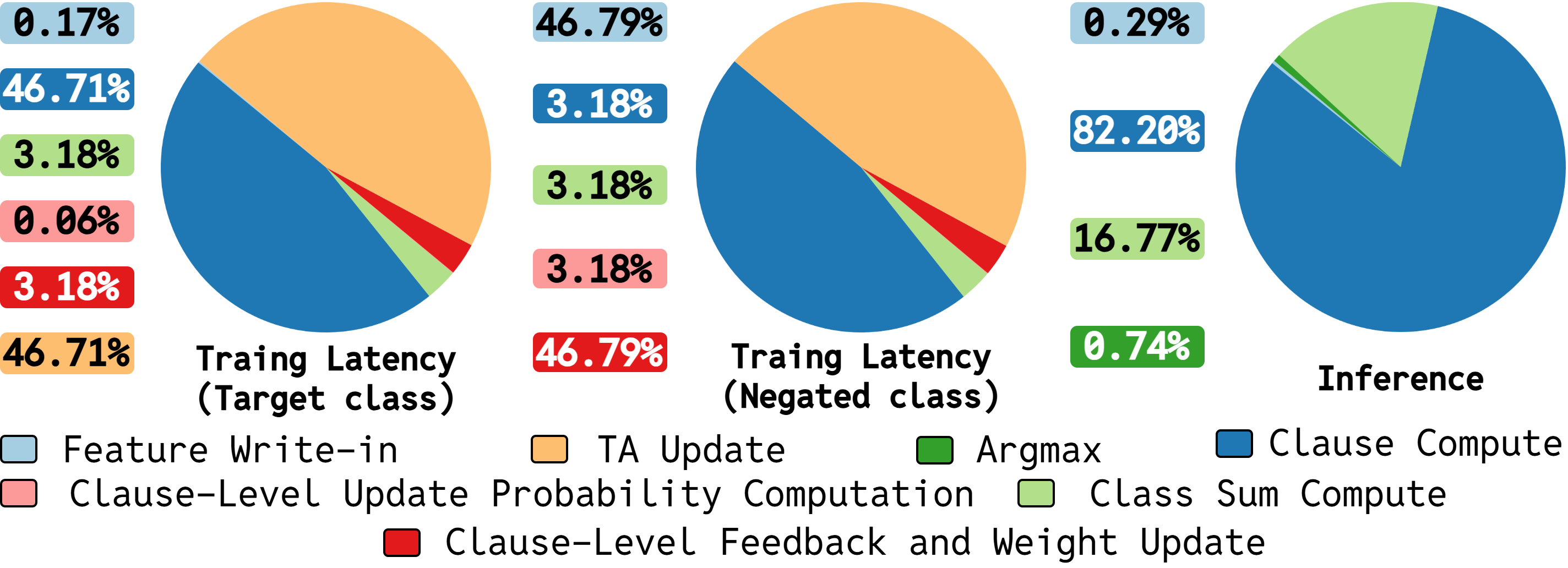}
    \vspace{-6mm}
    \caption{\small{Latency during training and inference for DTM-L.}}
    \label{fig:Latency}
\end{figure}
 
The flexibility of DTMs contrasts sharply with the fixed, customized designs seen in Conv-TM accelerator~\cite{Svein_Conv}, SATA~\cite{SATA}, FireFly~\cite{FireFly}, CNN~\cite{CNN_1}, SNN~\cite{SNN_1} implementations shown in Table~\ref{tab:comparison}. Although optimized for specific tasks, they lack the adaptability required for diverse applications or on-field model recalibration. The simplicity of the TM training process makes DTM less reliant on DSPs than the DNN accelerators. The DTM algorithms are mapped into LUTs and BRAMs except for the clause update probability comparison in clause feedback. This results in the trade-off between larger resource usage but lower power. 

In terms of training latency, DTM implementation demonstrates significantly lower latency compared to non-TM solutions SATA~\cite{SATA} and CNN~\cite{CNN_1}. The design in Conv TM~\cite{Svein_Conv} offers better training latency, but only for 128 clauses. This architecture requires all clauses in each patch to be computed in one clock cycle. Scaling up further would have a significant impact on resources because it stores all TA actions in the register buffer, limiting its scalability to larger models with more features or clauses required. This paper also provides preliminary inference insights for CIFAR-10 using Composite TM~\cite{grønningsæter2024optimizedtoolboxadvancedimage}. While they show the feasibility of TMs to solve more complex problems, with the addition of Convolution TM modules into the DTM framework, a DTM-based architecture would be able to offer better resource usage and \textit{flexibility}. Integration of these Conv TM modules is left to further work.

\begin{table*}[]
\centering
\caption{\small{Comparison with other FPGA accelerators suitable for similar edge sensor-based applications. (The ``-" indicates this result is not reported. The power in the brackets is the IP-only power for Conv TM and DTM designs. The training latency of DTM is calculated using the average training latency over 250 epochs. The training and inference latency rows show the values for MNIST, FMNIST and KMNIST respectively for the Conv-TM and DTM designs.})}

\label{tab:comparison}

\scriptsize
\vspace{-1mm}

\begin{tabular}{|c|c|c|c|c|c|c|c|}
\hline
Design                & SATA\cite{SATA}                & FireFly\cite{FireFly} & CNN\cite{CNN_1} & SNN\cite{SNN_1} & Conv TM\cite{Svein_Conv}             & DTM-L (Large)                         & DTM-S (Small)                        \\ \hline
Platform              & 65nm ASIC                                                                  & XCZU-3EG                                                        & XCZU-9EG                                                      & Kintex-7                                                      & XCZU-7EV                                                                          & XCZU-7EV                                                                           & XC7Z20                                                                            \\ \hline
Algorithm             & ANN,SNN                                                                    & SCNN                                                              & CNN                                                           & SNN                                                           & Conv TM                                                                           & \begin{tabular}[c]{@{}c@{}}Coalesced TM,\\ Vanilla TM\end{tabular}                 & \begin{tabular}[c]{@{}c@{}}Coalesced TM,\\ Vanilla TM\end{tabular}                \\ \hline
\begin{tabular}[c]{@{}c@{}}Accelerator\\Type\end{tabular}      & Training                                                                   & Inference                                                        & Training                                                      & Inference                                                     & Training                                                                          & Training                                                                           & Training                                                                          \\ \hline
Flexibility     & Reconfigurable                                                             & Reconfigurable                                                       & Customized                                                    & Customized                                                    & Customized                                                                        & Reconfigurable                                                                     & Reconfigurable                                                                    \\ \hline

LUT                   & \multirow{5}{*}{\begin{tabular}[c]{@{}c@{}}ASIC\\ Simulation\end{tabular}} & 15000                                                            & 32589                                                         & 46371                                                         & 196252                                                                            & 104222                                                                             & 43497                                                                             \\ \cline{1-1} \cline{3-8} 
FF                    &                                                                            & -                                                            & 33585                                                         & 30417                                                         & 73303                                                                             & 59610                                                                              & 33256                                                                             \\ \cline{1-1} \cline{3-8} 
DSP                   &                                                                            & 288                                                              & 143                                                           & 65                                                            & 129                                                                               & 25                                                                                 & 6                                                                                 \\ \cline{1-1} \cline{3-8} 
BRAM                  &                                                                            & 162                                                             & 95                                                            & 150                                                           & 0                                                                                 & 37                                                                                 & 138                                                                               \\ \cline{1-1} \cline{3-8} 
URAM                  &                                                                            & 0                                                                & 0                                                             & 0                                                             & 42.5                                                                              & 96                                                                                 & 0                                                                                 \\ \hline
Freq(MHz)             & 400                                                                        & 300                                                              & 100                                                           & 200                                                           & 50                                                                                & 100                                                                                & 50                                                                                \\ \hline
\begin{tabular}[c]{@{}c@{}}Test\\Accuracy\end{tabular}         & 99.00\% (MNIST)                                                                      & 98.12\% (MNIST)                                                         & 98.64\% (MNIST)                                                      & \begin{tabular}[c]{@{}c@{}}97.81\% (MNIST),\\ 83.16\% (FMNIST)\end{tabular}    & \begin{tabular}[c]{@{}c@{}}97.60\% (MNIST),\\ 84.10\% (FMNIST),\\ 82.80\% (KMNIST)\end{tabular}                & \begin{tabular}[c]{@{}c@{}}97.74\% (MNIST),\\ 86.38\% (FMNIST),\\ 83.11\% (KMNIST)\end{tabular}              & \begin{tabular}[c]{@{}c@{}}95.11\% (MNIST),\\ 80.62\% (FMNIST),\\ 72.52\% (KMNIST)\end{tabular}             \\ \hline
\begin{tabular}[c]{@{}c@{}}Training\\Latency(us)\end{tabular}  & 204.67                                                                     & Inference only                                                                & 3581.81                                                      & Inference only                                                             & 27, 27, 27                                                                                & 88.48, 97.69, 98.92                    & 50.08, 58.67, 60.57                   \\ \hline
\begin{tabular}[c]{@{}c@{}}Inference\\Latency(us)\end{tabular} & -                                                                          & 491.16                                                            & -                                                             & 4290, 4290                                                          & 9.4, 9.4, 9.4                                                                               & 44.72, 44.72, 44.72                                                                              & 22.32, 22.32, 22.32                                                                             \\ \hline
Power(W)              & -                                                                          & 2.550                                                            & -                                                             & 0.535                                                         & 4.543 (2.000)                                                                             & 4.359 (1.646)                                                                              & 1.687 (0.424)                                                                             \\ \hline
\end{tabular}

\end{table*}

\textbf{Point 2: Flexibility and Target Usage: }Table II highlights the flexibility of the DTM training architecture using the KWS6 dataset. The same DTM design can switch between the CoTM and the Vanilla TM with different clauses. The accuracy is comparable to Xilinx's FINN approach~\cite{FINN} and MATADOR~\cite{Matador} but DTM sacrifices throughput in favor of \textit{flexibility} to change TM algorithm type, number of clauses and train online. As shown in the table, higher accuracies are possible through higher clauses but at the cost of lower throughput. For KWS6, needing faster throughput means choosing CoTM while better performance means switching to Vanilla TM.\footnote{This analysis of throughput implications vs number of clauses is continued in Supplementary Material for more datasets. These datasets are also reflective of the type of target usage for this DTM implementation.}  

\begin{figure}[t]
    \centering
    \includegraphics[width = 0.95\linewidth]{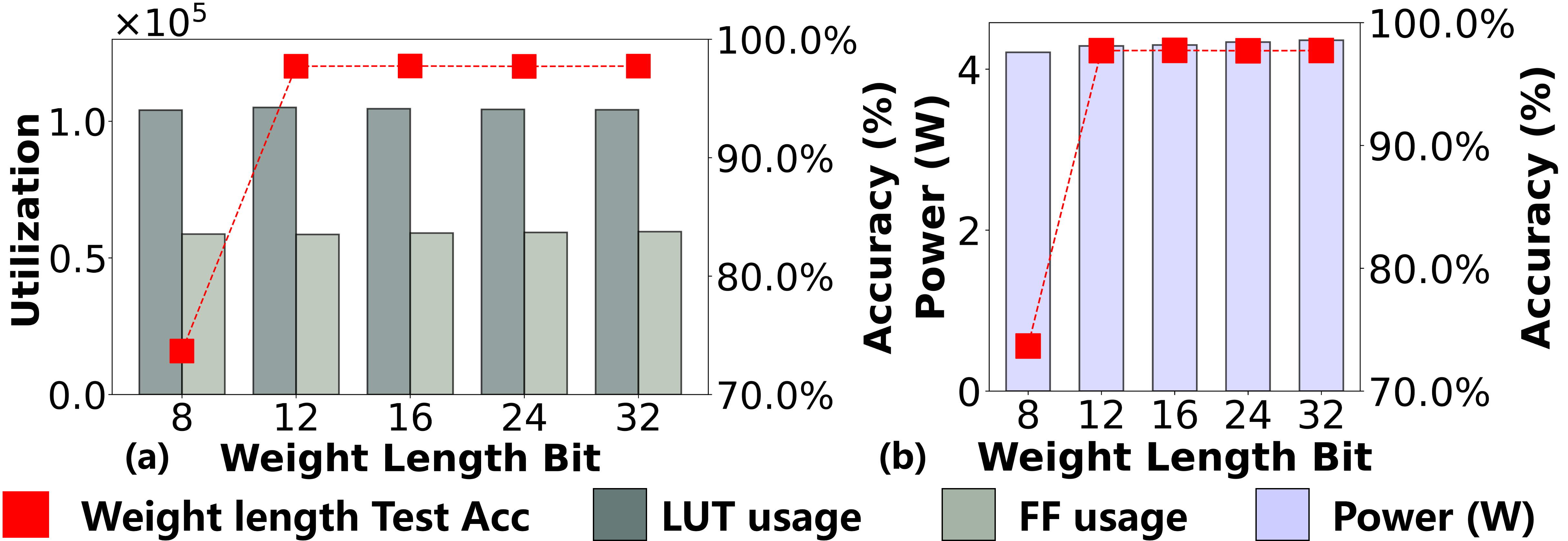}
    \vspace{-1mm}
    \caption{\small{Resource utilization and power for different weight length.}}
    \label{fig:Weight_Length}
\end{figure}
\begin{figure}[t]
    \centering
    \includegraphics[width = 0.95\linewidth]{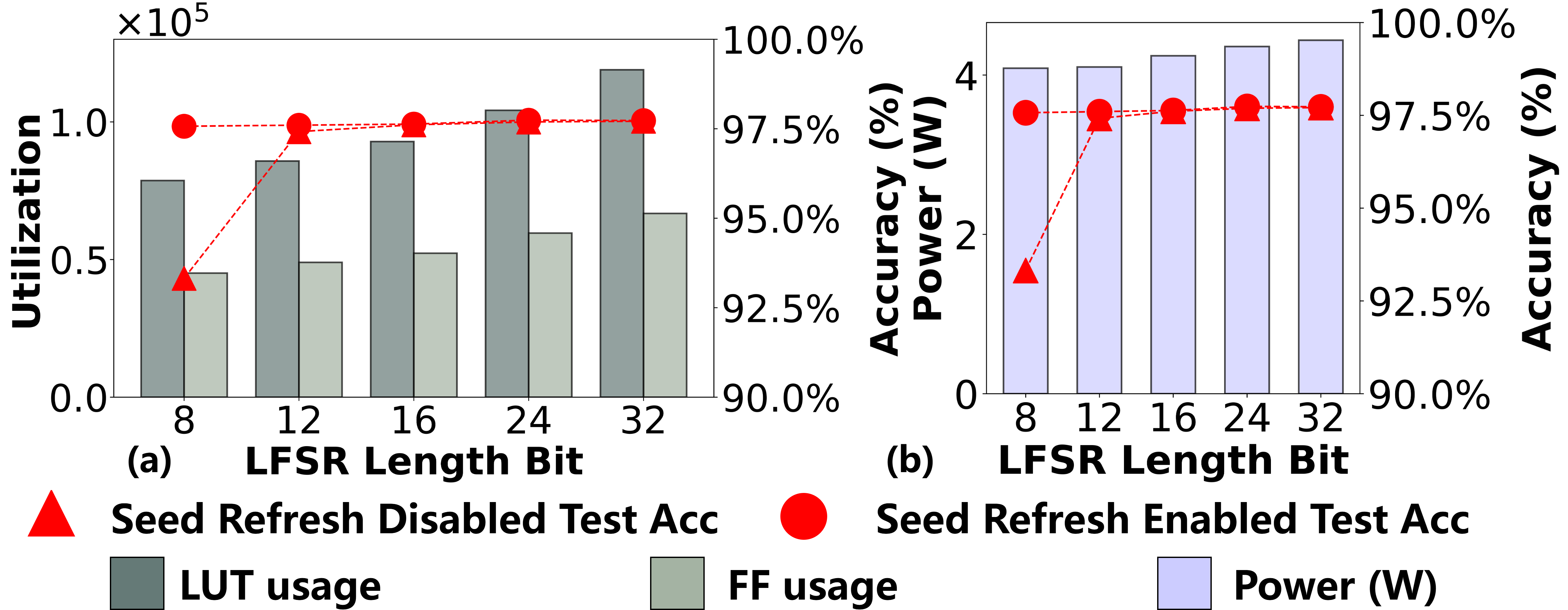}
    \vspace{-1mm}
    \caption{\small{Resource utilization and power for different LFSR length.}}
    \label{fig:LFSR_Length}
\end{figure}

\textbf{Point 3: Design Trade-offs: }The training process in the TM is affected by the quality of random numbers~\cite{10455073} and the precision of the CoTM weights. Figs.~\ref{fig:Weight_Length} and~\ref{fig:LFSR_Length} demonstrate the impact of learning efficacy (i.e, test accuracy) and the overhead in power and resources when varying the weight bit precision and LFSR length in the PRNG design. 

The results indicate that at least 12-bit weight precision is required to achieve good learning efficacy on the MNIST dataset (Fig.~\ref{fig:Weight_Length}a); however, the test accuracy, utilization, and power consumption saturate as weight precision increases, suggesting 12-bits are sufficient. Fig.~\ref{fig:LFSR_Length}a and Fig.~\ref{fig:LFSR_Length}b focus on learning efficacy, power, and resources when varying the LFSR length in the PRNG block. Higher precision and better quality random numbers yield better test accuracy, but the improvement becomes less substantial. Notice that with seed refreshing in the slave PRNG blocks, the accuracy at 8-bits is much higher than without (Fig.~\ref{fig:LFSR_Length}a), seed refreshing leads to better accuracies across all LFSR precisions. The LFSR precision plays a larger role in the overall resource and power than the weight precision.  


\begin{table*}[h]
\centering
\vspace{-2mm}
\caption{\small{Comparison against MATADOR~\cite{Matador}, FINN~\cite{FINN_R} and DNNBuilder~\cite{8587697} using KWS-6 dataset.}}
\vspace{-1mm}
\label{tab:date_set}
\scriptsize

\begin{tabular}{|c|c|c|c|c|c|c|c|c|c|}
\hline
Design                          & Model                  & LUT                     & FF                     & DSP                 & BRAM                & URAM                & Test Acc & Training Datapoint/s & Inference Datapoint/s \\ \hline
\multirow{3}{*}{DTM-CoTM}       & 2000 Clauses per class & \multirow{6}{*}{104222} & \multirow{6}{*}{59610} & \multirow{6}{*}{25} & \multirow{6}{*}{37} & \multirow{6}{*}{96} & 86.07\%  & 18281                 & 42878                  \\ \cline{2-2} \cline{8-10} 
                                & 1000 Clauses per class &                         &                        &                     &                     &                     & 83.76\%  & 34,128                & 83,084                 \\ \cline{2-2} \cline{8-10} 
                                & 500 Clauses per class  &                         &                        &                     &                     &                     & 81.15\%  & 63,696                & 163,241                \\ \cline{1-2} \cline{8-10} 
\multirow{3}{*}{DTM-Vanilla TM} & 700 Clauses per class  &                         &                        &                     &                     &                     & 87.12\%  & 45591                 & 25873                  \\ \cline{2-2} \cline{8-10} 
                                & 500 Clauses per class  &                         &                        &                     &                     &                     & 85.87\%  & 58,603                & 35,326                 \\ \cline{2-2} \cline{8-10} 
                                & 300 Clauses per class  &                         &                        &                     &                     &                     & 83.17\%  & 86,663                & 55,670                 \\ \hline
MATADOR                         & 700 Clauses per class  & 6063                    & 10658                  & 0                   & 3                   & 0                   & 87.10\%  & Inference only        & 8333333                \\ \hline
FINN                            & BNN (377-512-256-6)    & 42757                   & 45473                  & 0                   & 26.5                & 0                   & 84.60\%  & Inference only        & 750188                 \\ \hline
DNN Builder                     & DNN (273-64-128-64-6)  & 3035                    & 3790                   & 4                   & 36                  & 0                   & 81.00\%  & Inference only        & 5703                   \\ \hline
\end{tabular}

\end{table*}

\section{Conclusions and Future Work}

This paper presented an FPGA-based Dynamic TM (DTM) accelerator architecture that supports both inference and training acceleration for Vanilla TM and CoTMs. The architecture offers a competitive GOP/s/W ratio against comparable works but can also be reconfigured at run-time to adapt to different models and edge sensor tasks. This flexibility makes it suitable for on-field recalibration and personalized training. The TM clause computation maps very efficiently to FPGAs and the DTM design is parameterized to target both smaller low-power eFPGAs and larger FPGA fabrics. DTM is envisioned as a modular framework that can create a reconfigurable accelerator out of all TM algorithms. This paper explored the Vanilla and Coalesced variants. However, future work will extend to templates for Convolution and Regression along with Composite TM~\cite{granmo2023tmcompositeTM} support for larger datasets (CIFAR-10 and beyond). Users can use the FPGA to create required subsets from the global TM library for their application.


%

\appendices



\ifCLASSOPTIONcaptionsoff
  \newpage
\fi



%
\bibliographystyle{IEEEtran}
\bibliography{Bib}

%

\begin{IEEEbiography}
[{\includegraphics[width=1in,height=1.25in,clip,keepaspectratio]{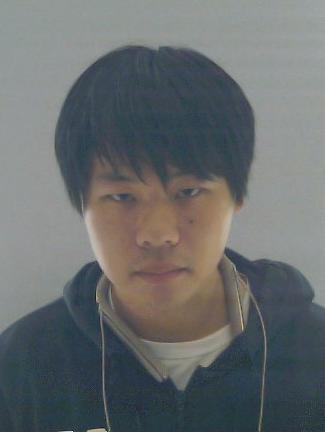}}]
{Gang Mao} received B.E degree from Northeastern University, China in 2016, and M.E degree from Newcastle University, Newcastle upon Tyne, UK. He is a PhD student at Newcastle University, Newcastle upon Tyne, UK. His research interests are focused on asynchronous circuit design and developing Machine Learning accelerators.
\end{IEEEbiography}

\begin{IEEEbiography}
[{\includegraphics[width=1in,height=1.25in,clip,keepaspectratio]{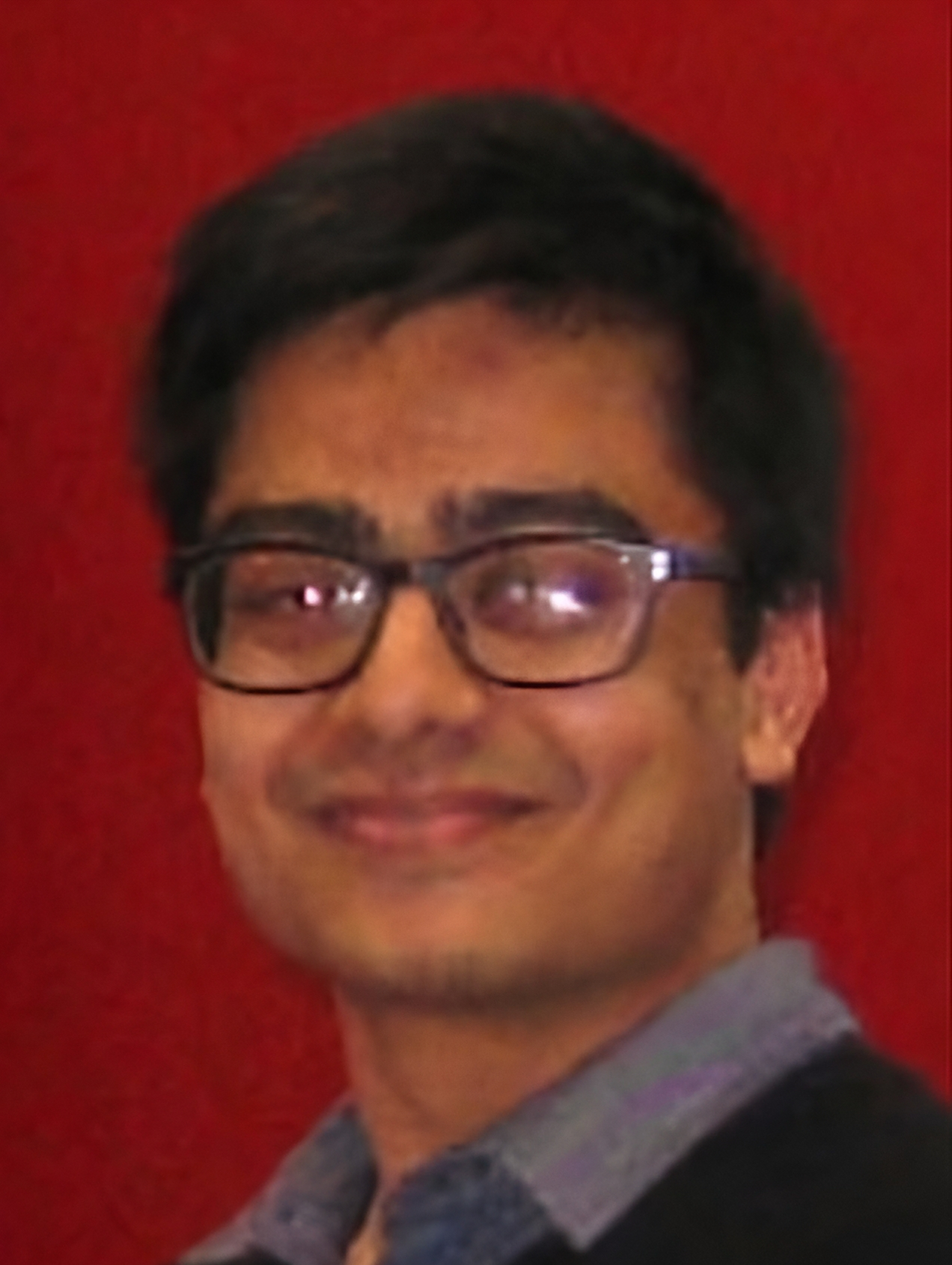}}]{Tousif Rahman} is a PhD student at Newcastle University, Newcastle upon Tyne, UK. He holds an MEng degree from the same institution. His research interests are focused on designing novel algorithms, architectures, and automation tools for translating Machine Learning applications to low-power, energy-efficient edge devices.  
\end{IEEEbiography}

\begin{IEEEbiography}
[{\includegraphics[width=1in,height=1.25in,clip,keepaspectratio]{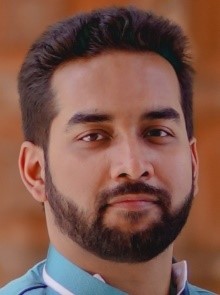}}]
{Sidharth Maheshwari} is an Assistant Professor in the department of Computer Science and Engineering at Indian Institute of Technology (IIT) Jammu. He completed his B.Tech in Electronics and Electrical Engineering at IIT Guwahati in 2013. He worked at Newcastle University for his PhD and Postdoc between 2016-2022. His interests include using hardware/software co-design to mitigate computational and energy bottlenecks of Big Data applications. He is looking towards solving real-world challenges such as improving water quality through molecular diagnostics that includes genome analysis and antimicrobial resistance surveillance. He is also working in the domain of battery-management systems and novel battery-pack design for Indian tropical climatic conditions in order to address challenges in the e-mobility sector. Another vertical in his research interest lies in exploring novel machine learning algorithms. 
\end{IEEEbiography}

\begin{IEEEbiography}
[{\includegraphics[width=1in,height=1.25in,clip,keepaspectratio]{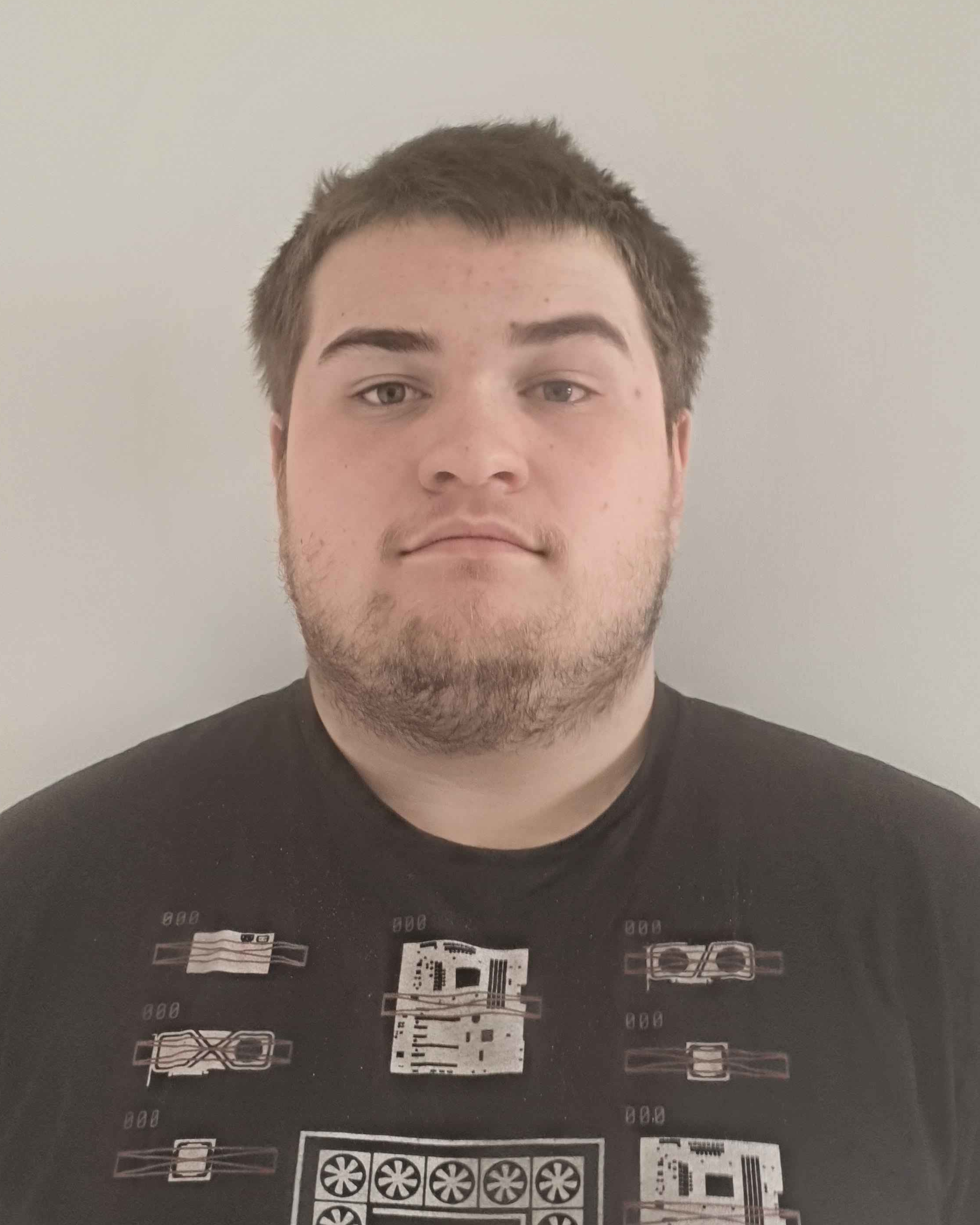}}]
{Bob Pattison} received a BEng (1st class Hons) degree from Newcastle University in 2024. Currently a 1st year PhD student in the Microsystems group at Newcastle University. His research focus is on explainable and energy-efficient machine learning algorithms for low-power devices. 
\end{IEEEbiography}

\begin{IEEEbiography}[{\includegraphics[width=1in,height=1.25in,clip,keepaspectratio]{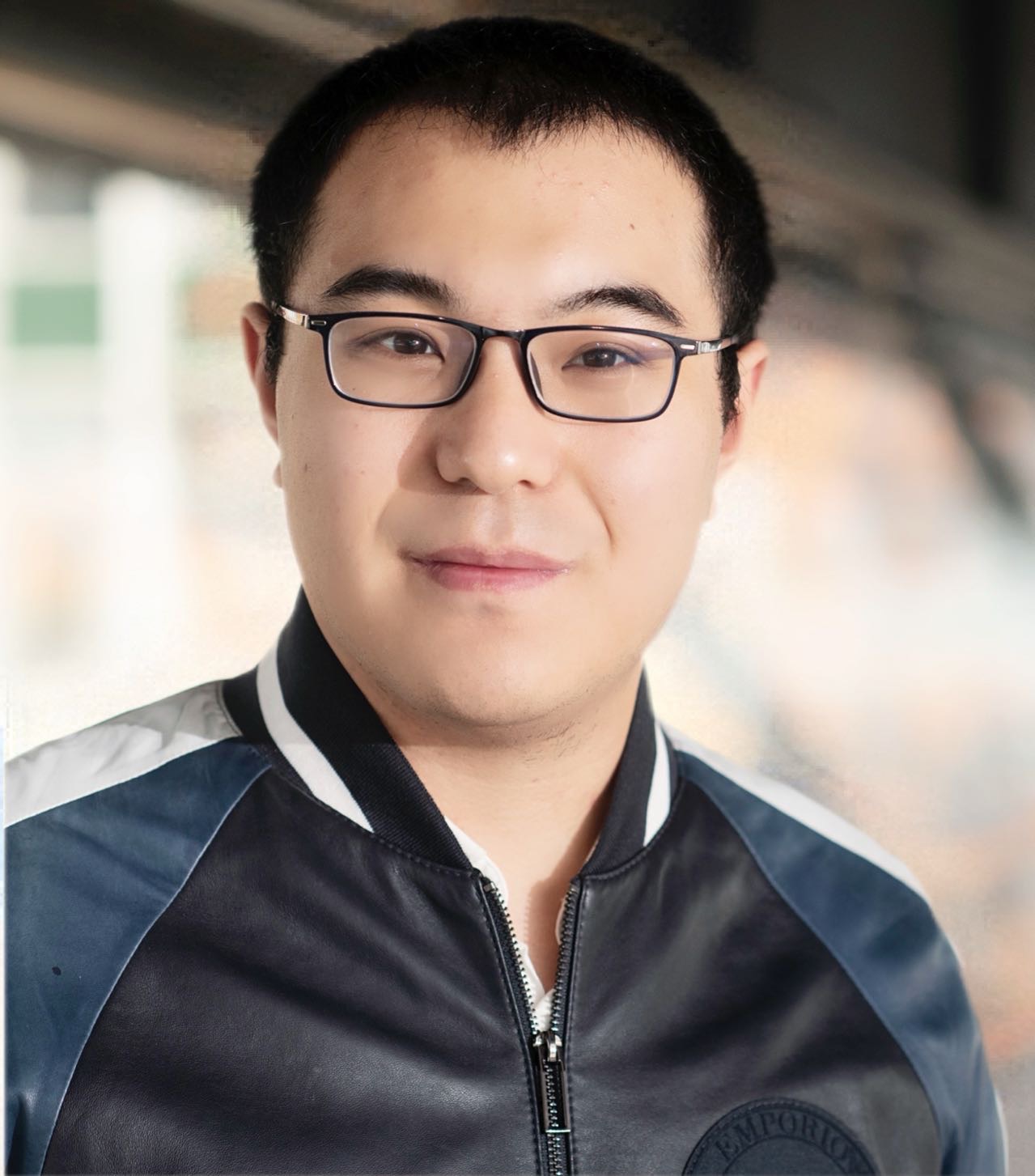}}]
{Zhuang Shao} is a Lecturer in Data Engineering and AI with the School of Engineering, Newcastle University, Newcastle upon Tyne, UK. He holds a PhD from the University of Warwick. His research interests include hardware-software co-design, low energy consumption machine learning, and vision-language learning. 
\end{IEEEbiography}

\begin{IEEEbiography}
[{\includegraphics[width=1in,height=1.25in,clip,keepaspectratio]{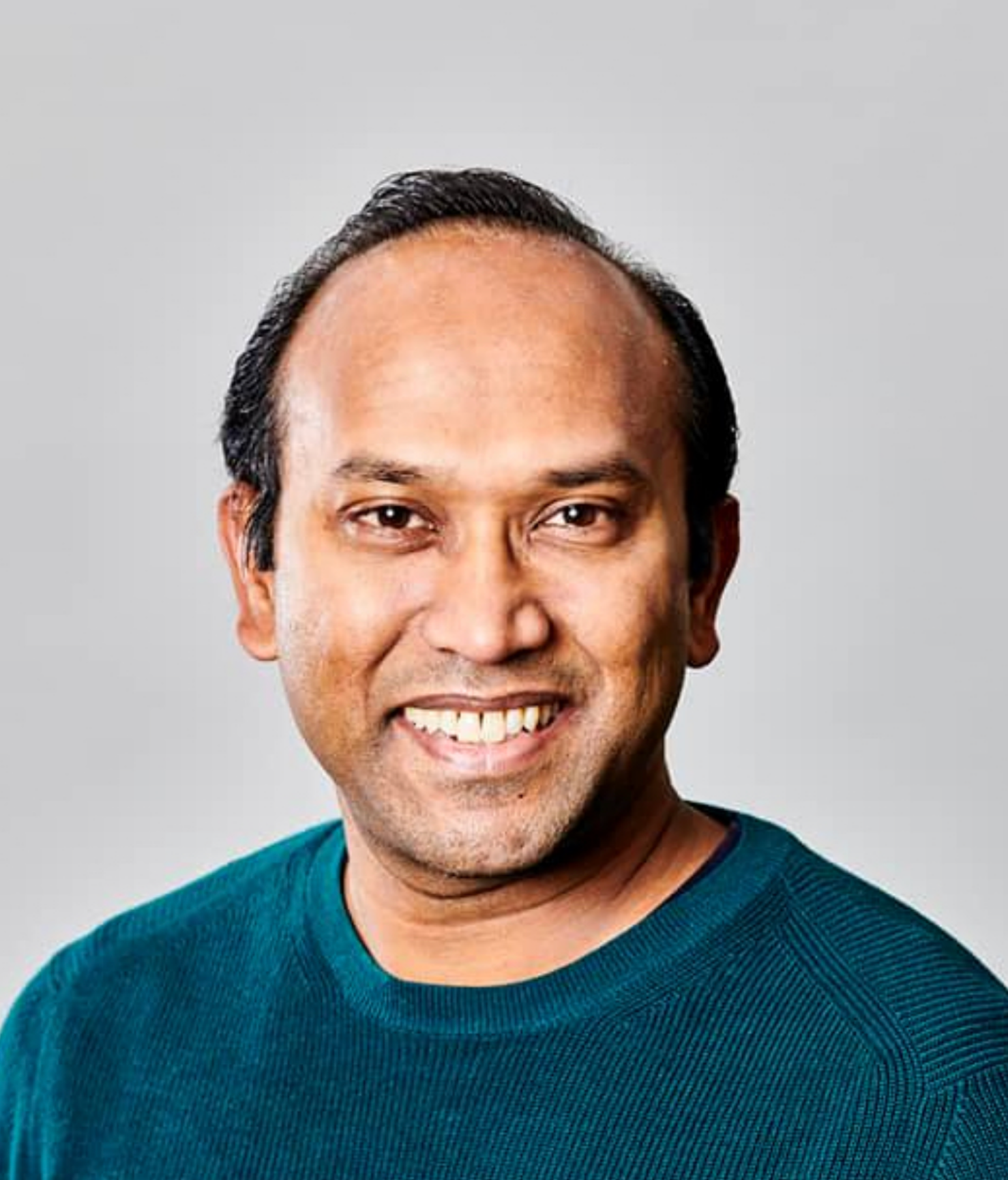}}]
{Rishad Shafik} is a Professor of Microelectronic Systems within the School of Engineering, Newcastle University, UK. Dr Shafik received his PhD, and MSc (with distinction) degrees from Southampton in 2010, and 2005; and BSc (with distinction) from the IUT, Bangladesh in 2001. He is one of the editors of the Springer USA book ``Energy-efficient Fault-tolerant Systems''. He is also author/co-author of 200+ IEEE/ACM peer-reviewed articles, with 4 best paper nominations and 4 best paper/poster awards. He recently chaired multiple international conferences/symposiums, UKCAS2020, ISTM2022; guest edited two special theme issues in Royal Society Philosophical Transactions A; he recently co-chaired 2nd International Symposium on the 
TM (ISTM), 2023. His research interests include hardware\slash software co-design for energy-efficiency and autonomy.
\end{IEEEbiography}

\begin{IEEEbiography}
[{\includegraphics[width=1in,height=1.25in,clip,keepaspectratio]{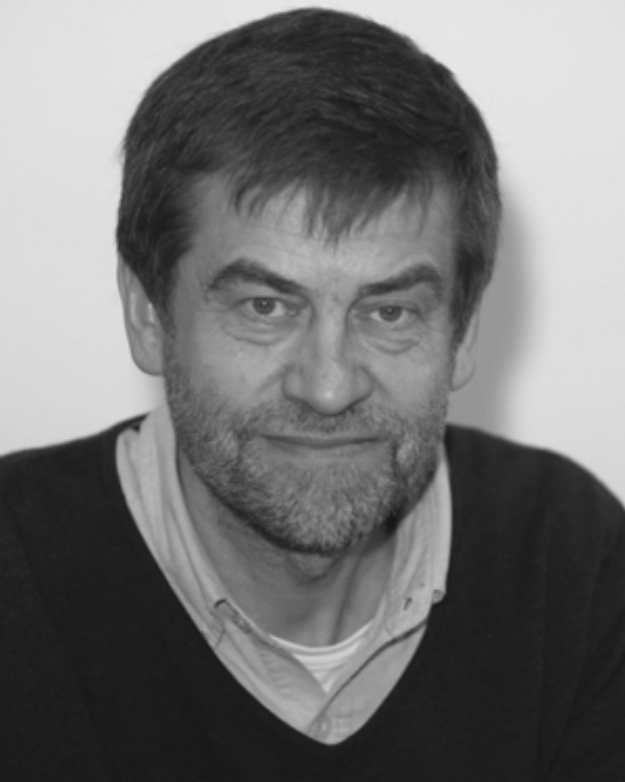}}]
{Alex Yakovlev} received the Ph.D. degree from the St. Petersburg Electrical Engineering Institute, St. Petersburg, USSR, in 1982, and D.Sc. from Newcastle University, UK, in 2006. He is currently a Professor of Computer Systems Design, who founded and leads the Microsystems Research Group, and co-founded the Asynchronous Systems Laboratory, Newcastle University. He was awarded an EPSRC Dream Fellowship from 2011 to 2013. He has published more than 500 articles in various journals and conferences, in the area of concurrent and asynchronous systems, with several best paper awards and nominations. He has chaired organizational committees of major international conferences. He has been principal investigator on more than 30 research grants and supervised over 70 Ph.D. students. He is a fellow of the Royal Academy of Engineering, UK.
\end{IEEEbiography}







\end{document}